\documentclass[%
reprint,
superscriptaddress,
amsmath,amssymb,
aps,
prb,
]{revtex4-2}

\usepackage{graphicx}
\usepackage{dcolumn}
\usepackage{bm}
\usepackage{hyperref}
\hypersetup{hypertex=true, colorlinks=true, linkcolor=blue, anchorcolor=blue, citecolor=blue, urlcolor=blue}
\usepackage[mathlines]{lineno}
\usepackage{multirow}

\begin{document}
	
\preprint{APS/123-QED}
	
\title{Nontrivial topology in one- and two-dimensional asymmetric systems with chiral boundary states}
	
\author{Yunlin Li}
\affiliation{%
	Department of Illuminating Engineering and Light Sources, College of Intelligent Robotics and Advanced Manufacturing, Fudan University, Shanghai 200433, China}%
	
\author{Yufu Liu}
\affiliation{%
	Department of Illuminating Engineering and Light Sources, College of Intelligent Robotics and Advanced Manufacturing, Fudan University, Shanghai 200433, China}%
		
\author{Xuezhi Wang}%
\affiliation{%
	Department of Illuminating Engineering and Light Sources, College of Intelligent Robotics and Advanced Manufacturing, Fudan University, Shanghai 200433, China}%
		
\author{Haoran Zhang}
\affiliation{%
	Department of Illuminating Engineering and Light Sources, College of Intelligent Robotics and Advanced Manufacturing, Fudan University, Shanghai 200433, China}%
	
\author{Xunya Jiang}%
\email{jiangxunya@fudan.edu.cn}
\affiliation{%
	Department of Illuminating Engineering and Light Sources, College of Intelligent Robotics and Advanced Manufacturing, Fudan University, Shanghai 200433, China}%
	
\begin{abstract}
	
Symmetry plays an important role in the topological band theory. In contrary, study on the topological properties of the asymmetric systems is rather limited, especially in higher-dimensional systems. In this work, we explore a new theory to study the topology in various one-dimensional (1D) and two-dimensional (2D) asymmetric systems with chiral boundary states. Starting from the simple SSH$m$ model, we show the chiral topology of its edge states by redefining sublattices. Meanwhile, based on its Rice-Mele-like effective Hamiltonian, a new topological invariant $\bar{Z}$ can be defined and the bulk-edge correspondence is established. With this clear physical picture, our theory can be extended to the more complex asymmetric ladder models, or even the 2D asymmetric systems. In the 2D BBH$3$ model, new chiral corner states with redefined lattices are found based on our method. These corner states are independent of any spatial symmetry and exhibit the characteristics of topological bound states in the continuum (TBICs). Moreover, the topological invariant can be calculated by introducing $\bar{Z}$ into 2D. At last, we propose an acoustic experiment of the BBH3 model where chiral corner states are numerically observed. Our work exhibits a new approach to study the topological properties of asymmetric systems. By redefining sublattices, we find that the models with entirely different structures might share the same topological origins.
		
\end{abstract}
	
\maketitle
	
	
\section{\label{sec:level1}INTRODUCTION}
In the last few decades, topological insulator has attracted much attention in both quantum and classic systems \cite{RevModPhys.83.1057,RevModPhys.88.021004,RevModPhys.91.015006,kim2020recent}. Since the discovery of spin-orbit-induced topological insulators \cite{bernevig2006quantum,konig2007quantum}, symmetry-protected topological (SPT) phases of matter play an important role in the study of topological band theory. In nontrivial SPT phases, symmetry-protected topological edge states can be found for the systems under open boundary condition (OBC) and the bulk-edge correspondence can be established. For instance, in one-dimensional (1D) non-superconducting systems, based on the tenfold classification in topological insulators and superconductors \cite{PhysRevB.78.195125,RevModPhys.88.035005}, nontrivial topological phases protected by chiral symmetry can be found. Moreover, for 1D systems with spatial inversion symmetry, Abelian topological phases \cite{PhysRevB.83.245132,PhysRevX.4.021017,PhysRevB.96.245115} described by Zak phases \cite{PhysRevLett.62.2747} and non-Abelian topological phases \cite{guo2021experimental,jiang2021four} described by quaternion groups are extensively studied. As to the two-dimensional (2D) systems, with the development of higher-order topological insulators \cite{ni2019observation,xie2021higher}, topological corner states protected by spatial symmetry \cite{PhysRevB.96.245115,PhysRevB.99.245151} or chiral symmetry \cite{PhysRevLett.125.166801,PhysRevLett.128.127601} are found theoretically and experimentally \cite{doi:10.1126/science.aah6442,PhysRevLett.118.076803,PhysRevLett.122.233902,PhysRevLett.122.233903,PhysRevLett.124.206601,he2020quadrupole,PhysRevLett.131.157201,PhysRevApplied.20.064042,li2025Z}.

Recently, studies have shown that rich topological phenomena can also be observed in many 1D asymmetric systems \cite{PhysRevA.99.013833,PhysRevLett.128.093901,wang2023sub,PhysRevB.110.035106,k77w-ft26,PhysRevB.106.085109,marques2020analytical,guo2022rotation,du2024one,PhysRevB.110.125424,Yan:23,PhysRevB.110.174311}. In particular, the topology protected by ``hidden symmetries" has attracted much attention. One example is the sub-symmetry \cite{wang2023sub,PhysRevResearch.6.033323} which is the chiral symmetry defined on certain sublattices and can protect nontrivial topology and chiral edge states in the presence of certain symmetry-breaking terms. Another example is the latent symmetry \cite{PhysRevB.110.035106,PhysRevLett.126.180601}, which is defined by effective Hamiltonian through isospectral reduction. When the effective Hamiltonian is mirror symmetric, the nontrivial topology described by the quantized Zak phases can be observed. Meanwhile, new theoretical methods have been utilized to study the topology of asymmetric systems. A typical example is the non-Bloch theory method \cite{k77w-ft26,PhysRevB.110.035424} which is based on the solution of periodic systems in complex-$k$-$\omega$ space. Very recently, the homomorphic mapping method to reconstruct the nontrivial topology for Rice-Mele-like systems is found and can be extended to the systems with asymmetric perturbations \cite{PhysRevB.110.104106}.

These inspiring methods shed a light on the rich topological properties in asymmetric systems. However, so far these methods all have their own limitations. For example, for systems with sub-symmetry the topological invariant is hard to define \cite{wang2023sub}, for systems with latent symmetry the specifically designed models from graph theory \cite{PhysRevLett.126.180601} are required, while non-Bloch theory has difficulties to be extended to 2D systems \cite{PhysRevLett.123.066404}. These limitations hinder the study of corresponding topological phenomena. Therefore, to develop a more general theory to illustrate the topological properties of different asymmetric systems with clear physical picture is an urgent need. With such theory, a topological invariant can be well defined and new topological edge states or corner states can be predicted for asymmetric systems. Even more, since its clear and simple physical picture, such theory can be extended to 2D or higher-dimensional systems.

\begin{figure}[t]
	\centering
	\includegraphics[width=1\linewidth]{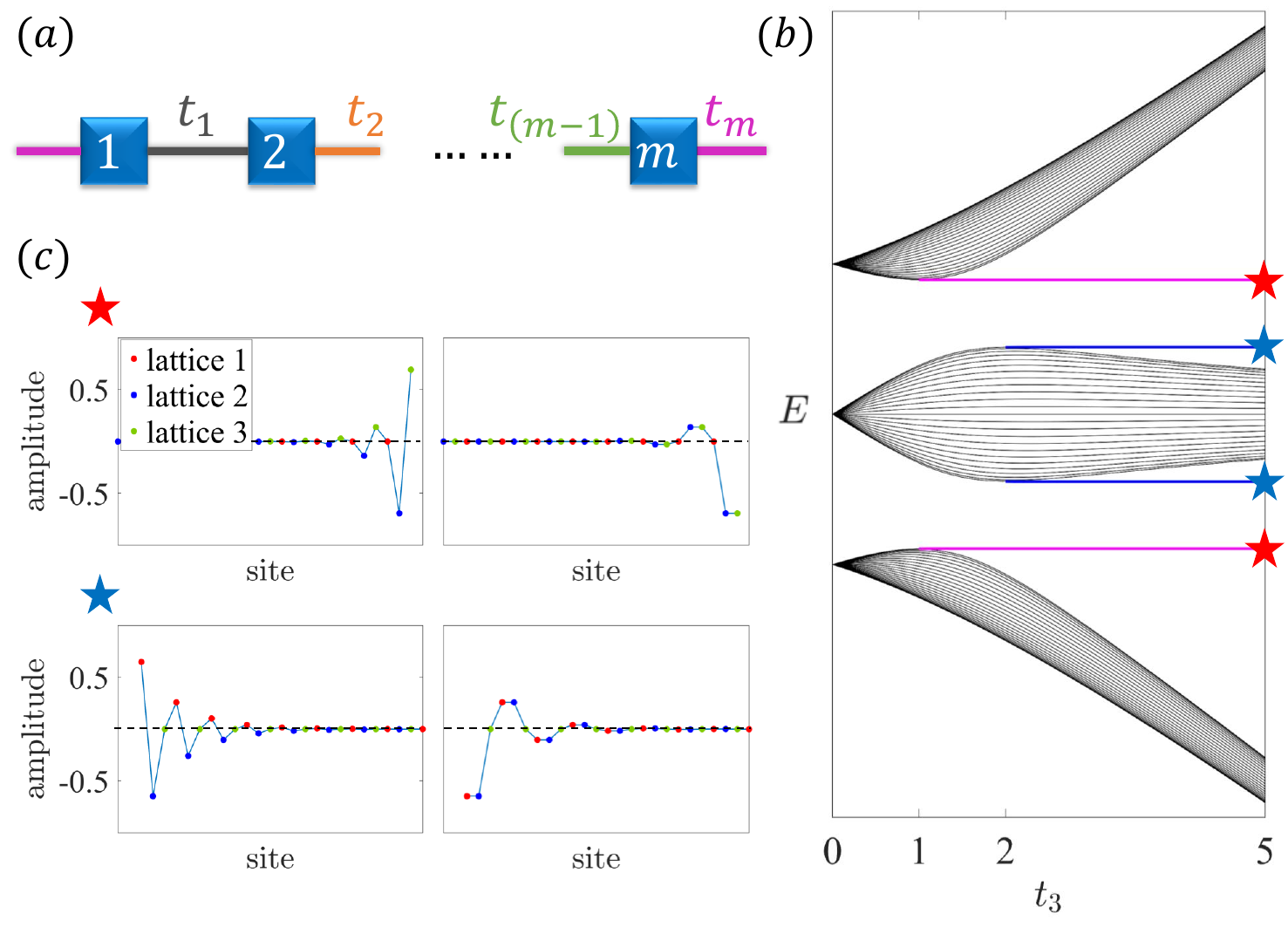}
	\caption{ (a) Schematic of SSH$m$ model. (b) Band structure versus $t_3$ under OBC with $t_1=1$, $t_2=2$. The edge states located at the left (right) boundary are marked by blue (red) lines. (c) The spatial distribution of the edge states marked in (b). } \label{fig:1}
\end{figure}

In this work, by redefining sublattices, we develop a new theory to study the topology for both 1D and 2D asymmetric systems where new topological chiral edge states (or corner states) are predicted and the corresponding bulk-edge (edge-corner) correspondence can be deduced. First, we introduce our theory step by step with a class of the simplest 1D asymmetric chain models, namely the SSH$m$ model. Based on the non-Bloch theory, we analytically study the wavefunctions of the possible edge states for the model and show that, under redefined sublattices, all of them can be understood as topological chiral edge states protected by sub-symmetry. With the redefined lattices, we also predict the breaking of bulk-edge correspondence in the extended SSH$3$ model with inter-cell long-range hoppings. As to the bulk-edge correspondence, we find that the chiral edge states originate from the Rice-Mele-like effective Hamiltonian. Based on the reduced wavefunction of the effective Hamiltonian, a new quantized topological invariant $\overline{Z}$ can be defined through normalization. With this clear physical picture, our theory can be extended to more complex systems. As an example, we establish the bulk-edge correspondence in the asymmetric ladder SSH model. To take a step further, we introduce our theory into 2D systems. We construct the 2D BBH$3$ model where new topological chiral corner states with redefined sublattices are found. These corner states are independent of spatial symmetries and fundamentally different from the conventional topological quadrupole corner states. The corner states can even move into the bulk continuous spectrum, exhibiting a new kind of the topological bound states in the continuum (TBICs). In terms of the topological invariant, we introduce $\overline{Z}$ into 2D systems and establish the edge-corner correspondence through the quantized normalized edge polarization $\bar{p}^{edge}$. At last, our theory can be easily extended to other systems, such as the acoustic or photonic systems. We propose a laboratory acoustic experiment to test new topological phenomena of the asymmetric BBH$3$ model where the new chiral corner states could be numerically observed. Our work discovers the topological chiral characteristics of the boundary states in various asymmetric systems. Moreover, by redefining sublattices, we show that asymmetric models with entirely different structures might share the same topological origins while the bulk-edge correspondence can be established with a general topological invariant.

\section{\label{sec:level1}TOPOLOGICAL CHIRAL EDGE STATES IN SSH\boldmath{$m$} MODEL} \label{II}

We would like to start with a class of the simplest 1D asymmetric models, i.e. the SSH$m$ model. As a natural extension of the well-known SSH model \cite{PhysRevLett.42.1698}, the SSH$m$ model is the 1D chain model whose unit cell contains $m$ sublattices \cite{marques2020analytical}. As shown in Fig. \ref{fig:1}(a), the Hamiltonian of the SSH$m$ model can be written as:
\begin{eqnarray} \label{eq1}
	H_m=\sum_{n}[\sum_{j=1}^{m-1}t_jc_{n,(j+1)}c^{\dagger}_{n,j}+t_mc_{(n+1),1}c^{\dagger}_{n,m}+h.c.] \notag \\
\end{eqnarray}
where $n$ is the index of the unit cell, $j$ is the index of the sublattice. Here we only consider the nearest-neighbour hoppings. When $t_i \neq t_j$ and $m$ is odd, both spatial inversion symmetry and chiral symmetry are broken. For the case $m=3$, the band structure and the edge state wavefuntions of the asymmetric SSH3 model \cite{PhysRevApplied.23.L031001,PhysRevB.106.085109,PhysRevB.110.174311} are shown in Fig. \ref{fig:1}(b) and \ref{fig:1}(c), respectively. Two critical phenomena can be observed. Firstly, the emergence of the edge state does not coincide with the gap closing \cite{PhysRevB.106.085109}, indicating that the conventional topological phase transition cannot happen in this model. Secondly, for the wavefunctions of the edge states, the amplitude at certain sublattice remains zero \cite{PhysRevB.110.125424}, which shares some similarity with the characteristics of the edge states protected by chiral symmetry \cite{wang2023sub}.
	
\subsection{\label{sec:level2}Non-Bloch Theory for SSH\boldmath{$m$} Model}

In this section, we would like to study the edge states of the SSH$m$ model utilizing the non-Bloch theory \cite{PhysRevB.110.035424,verma2024topological}. For the SSH$m$ model with $N$ unit cells under OBC, its eigenstates take the form $| \Psi \rangle = 1/A\sum_{n=1}^{N}\sum_{j=1}^{m}\psi_{n,j}| n,j \rangle$ with $\psi_{n,j} \equiv \psi_{n,j}(z_1,z_2) = c_1\phi_j^{(1)}z_1^n + c_2\phi_j^{(2)}z_2^n$ where $A$ is the normalization coefficient, $z_\alpha \in C$ is the generalized complex momentum with $\alpha = 1,2$. $| \Psi \rangle$ should satisfy the bulk condition, i.e. $\{ z_\alpha , \phi_j^{(\alpha)} \}$ are the solutions of the following bulk equation with a given $E$:
\begin{eqnarray} \label{eq2}
	\mathcal{H}_m(z)\begin{bmatrix}
		\phi_1 \\
		\phi_2 \\
		...\\
		\phi_m
	\end{bmatrix} = E\begin{bmatrix}
	\phi_1 \\
	\phi_2 \\
	...\\
	\phi_m
	\end{bmatrix},
\end{eqnarray}
where $\mathcal{H}_m(z)$ can be obtained by applying the substitution $e^{ik} \rightarrow z$ to the momentum-space Hamiltonian $\mathcal{H}_m(k)$ of the SSH$m$ model:
\begin{eqnarray} \label{eq8}
	\mathcal{H}_m(z) = \begin{bmatrix}
		0 & t_1\quad & & & & & t_m/z \\
		t_1 & 0 & t_2 & & & &  \\
		 & t_2 & 0 & t_3 & & &\\
		 & & \ddots & \ddots & \ddots & & \\
		 & & & t_{m-3} & 0 & t_{m-2} & \\
		 & & & & t_{m-2} & 0 & t_{m-1}\\
		t_mz & & & & & t_{m-1} & 0
	\end{bmatrix}.
\end{eqnarray}
From Eq. \eqref{eq2}, we have $z_1z_2=1$. Without loss of generality, let $|z_1| \leq |z_2|$. On the other hand, $| \Psi \rangle$ should also satisfy the boundary condition:
\begin{eqnarray} \label{eq3}
	M_b[c_1,c_2]^T=0
\end{eqnarray}
with
\begin{eqnarray} \label{eq63}
	M_b = \begin{bmatrix}
		P(z_1) & P(z_2) \\
		z_1^{N+1}Q(z_1) & z_2^{N+1}Q(z_2)
	\end{bmatrix}
\end{eqnarray}
where $P(z_\alpha) = -t_m\phi_m^{(\alpha)}$, $Q(z_\alpha) = -t_m\phi_1^{(\alpha)}$. When $\mathrm{det}[M_b]=0$, the nontrivial solution of Eq. \eqref{eq3} are exactly the eigenstates of the SSH$m$ model under OBC.

For $| \Psi \rangle$ representing the localized edge states, we have $|z_1| < 1 < |z_2|$. Together with $z_1^{N+1} \rightarrow 0$, one can obtain:
\begin{eqnarray} \label{eq4}
	\mathrm{det}[M_b]=-z_2^{N+1}P(z_1)Q(z_2)=0.
\end{eqnarray}
Two solutions can be obtained from Eq. \eqref{eq4}, i.e. $P(z_1) = 0$ and $Q(z_2) = 0$.

For $P(z_1) = 0$, it follows that $\phi_m^{(1)} = 0$. Substituting into Eq. \eqref{eq2} yields:
\begin{eqnarray} \label{eq5}
	\left\{\begin{matrix}
		\mathcal{H}_{m-1}^L[\phi_1^{(1)},...,\phi_{m-1}^{(1)}]^T = E[\phi_1^{(1)},...,\phi_{m-1}^{(1)}]^T
		\\\\
		t_m\phi_1^{(1)}z_1 + t_{m-1}\phi_{m-1}^{(1)} = 0
	\end{matrix}\right.,
\end{eqnarray}
where
\begin{eqnarray} \label{eq6}
	\mathcal{H}_{m-1}^L = \begin{bmatrix}
		0 & t_1 & & & \text{{\huge 0}} \\
		t_1 & 0 & t_2 & & \\
		& \ddots & \ddots & \ddots\\
		& & t_{m-3} & 0 & t_{m-2}\\
		\text{{\huge 0}} & & & t_{m-2} & 0
	\end{bmatrix}
\end{eqnarray}
is the Hamiltonian of the subspace composed of the sublattices $1$ to $m-1$. Moreover, Eq. \eqref{eq5} indicates that $[\phi_1^{(1)},...,\phi_{m-1}^{(1)}]^T$ is the eigenstate of $\mathcal{H}_{m-1}^L$. When $|z_1| = |{t_{m-1}\phi_{m-1}^{(1)}}/{t_m\phi_1^{(1)}}| < 1$, $| \Psi \rangle$ stands for the edge estate localized at the left boundary of the system:
\begin{eqnarray} \label{eq7}
	| L \rangle = A_L\sum_{n=1}^{N}z_1^{n-1}\sum_{j=1}^{m}\phi_j^{(1)}| n,j \rangle.
\end{eqnarray}
Since $\mathcal{H}_{sub-1}$ is $(m-1)$-dimensional matrix, there are $m-1$ different eigenstates. Therefore, SSH$m$ model has at most $m-1$ left-boundary edge states.

Similarly, for $Q(z_2) = 0$, it follows that $\phi_1^{(2)} = 0$, which yields:
\begin{eqnarray} \label{eq9}
	\left\{\begin{matrix}
		\mathcal{H}_{m-1}^R[\phi_2^{(2)},...,\phi_{m}^{(2)}]^T = E[\phi_2^{(2)},...,\phi_{m}^{(2)}]^T
		\\\\
		t_m\phi_m^{(2)} + t_{1}\phi_{2}^{(2)}z_2 = 0
	\end{matrix}\right.,
\end{eqnarray}
where $\mathcal{H}_{m-1}^R$ is the Hamiltonian of the subspace composed of the sublattices $2$ to $m$. When $|z_2|=|t_m\phi_m^{(2)}/{t_1\phi_2^{(2)}}|>1$, $| \Psi \rangle$ stands for the edge estate localized at the right boundary of the system:
\begin{eqnarray} \label{eq10}
	| R \rangle = A_R\sum_{n=1}^{N}z_2^{n-1}\sum_{j=1}^{m}\phi_j^{(2)}| n,j \rangle.
\end{eqnarray}
Since $\mathcal{H}_{m-1}^R$ is also $(m-1)$-dimensional, at most $m-1$ right-boundary edge states can be obtained. In summary, all the possible edge states of the SSH$m$ model are given by Eq. \eqref{eq7} and Eq. \eqref{eq10}. It is worth noting that similar methods can also be utilized to analyze the topological chiral edge states in the disordered SSH model \cite{note}.

\subsection{\label{sec:level2}Sub-Symmetry Protected Chiral Edge States with Redefined Sublattices}

\begin{figure}[t]
	\centering
	\includegraphics[width=1\linewidth]{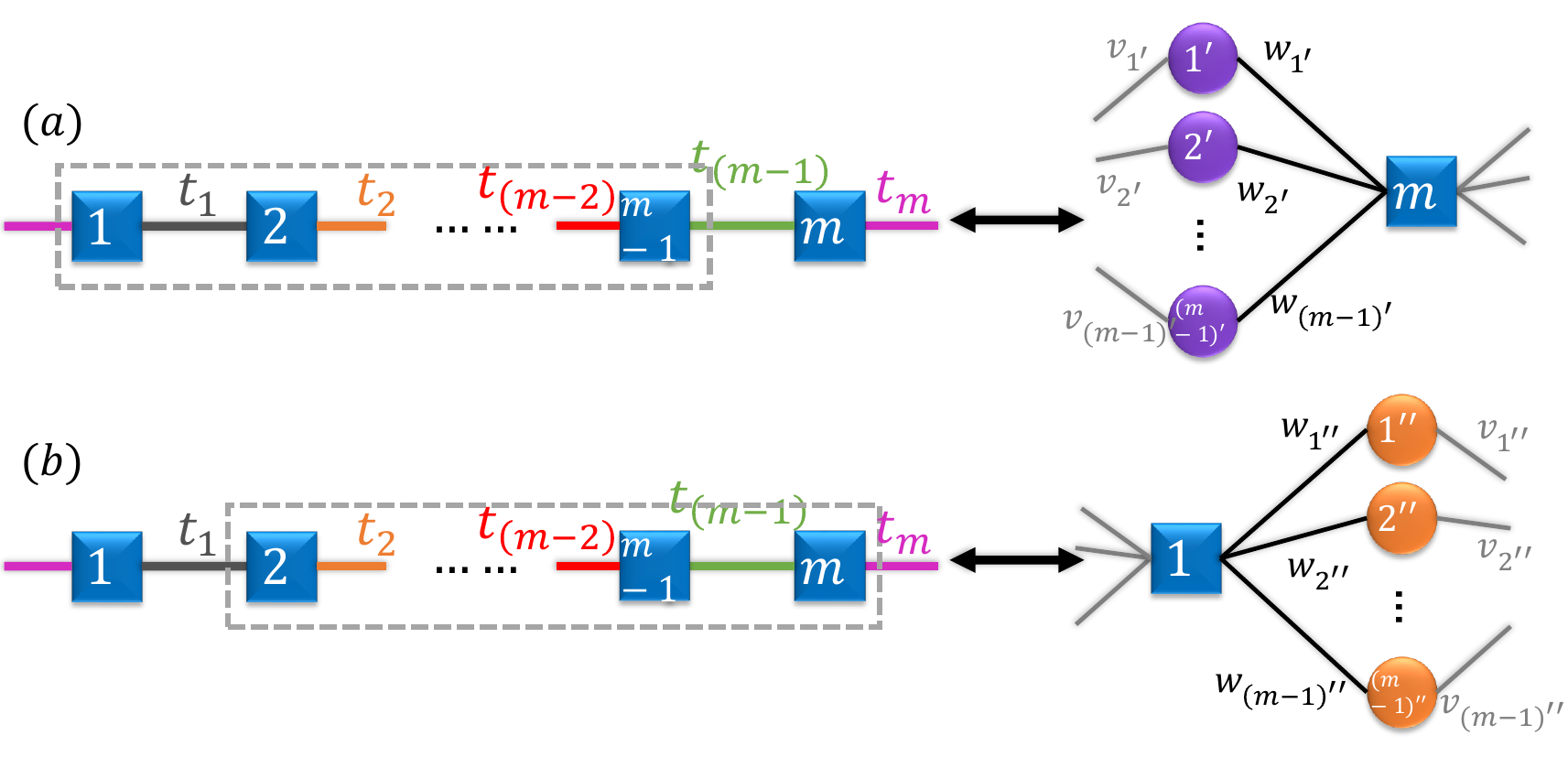}
	\caption{ (a)-(b) Schematic of the SSH$m$ model with redefined sublattices diagonalizing the subspace (a) $\mathcal{H}_{m-1}^L$; (b) $\mathcal{H}_{m-1}^R$ marked by dashed box. } \label{fig:2}
\end{figure}

In this section, we would like to study the topological origin of the edge states obtained in the last section. We first consider the edge states at the left boundary, i.e. the $| L \rangle$ in Eq. \eqref{eq7}. The form of $| L \rangle$ indicates that the left-boundary edge states are composed of the eigenstates of $\mathcal{H}_{m-1}^L$, which is the Hamiltonian of the subspace marked by dashed line in Fig. \ref{fig:2}(a). Therefore, we perform a proper representation transformation $U$ to diagonalize $\mathcal{H}_{m-1}^L$:
\begin{eqnarray} \label{eq64}
	\mathcal{H}_{m-1}^{L'} = U\mathcal{H}_{m-1}^{L}U^{\dagger} = \begin{bmatrix}
		V_{1'} & & & \text{{\huge 0}} \\
		 & V_{2'} & & \\
		&  & \ddots & \\
        \text{{\huge 0}} & & & V_{(m-1)'}
	\end{bmatrix}.
\end{eqnarray}
In this representation, the sublattices are redefined as $\{ | j' \rangle \}$, $j=1,...,m-1$. Then the Hamiltonian of the SSH$m$ model can be written as:
\begin{eqnarray} \label{eq65}
	\mathcal{H}_{m}'(k) = \begin{bmatrix}
		V_{1'} & & & \text{{\huge 0}} & t_{1'}(k)\\
		& V_{2'} & & & t_{2'}(k)\\
		&  & \ddots & & \vdots\\
		\text{{\huge 0}} & & & V_{(m-1)'} & t_{(m-1)'}(k)\\
		t_{1'}^{\dagger}(k) & t_{2'}^{\dagger}(k) & \cdots & t_{(m-1)'}^{\dagger}(k) & 0
	\end{bmatrix}, \notag \\
\end{eqnarray}
where $t_{j'}(k)=w_{j'}+v_{j'}e^{ik}$ and $w_{j'}$ and $v_{j'}$ can be calculated by:
\begin{eqnarray} \label{eq13}
	\left\{\begin{matrix}
		w_{j'} = \langle n,j' | H_{m} | n,m \rangle = t_{m-1}\phi_{m-1}^{j'}
		\\\\
		v_{j'} = \langle n,m | H_{m} | n+1,j' \rangle = t_m\phi_{1}^{j'} 
	\end{matrix}\right..
\end{eqnarray}
Here $\phi_i^{j'}$ is the component of the redefined $| j' \rangle$ on the original sublattice $| i \rangle$ and $H_{m}$ stands for the Hamiltonian of SSH$m$ model in Eq. \eqref{eq1}. As shown in Fig. \ref{fig:2}(a), $\mathcal{H}_{m}'(k)$ is equivalent to $m-1$ SSH models:
\begin{eqnarray} \label{eq11}
	H_m' = \sum_{j'=1'}^{(m-1)'}H_{j'}^{SSH} + V_{j'}P_{j'},
\end{eqnarray}
where
\begin{eqnarray} \label{eq12}
	H_{j'}^{SSH} = \sum_{n}w_{j'}c_{n,m}c^{\dagger}_{n,j'} + v_{j'}c_{(n+1),j'}c^{\dagger}_{n,m} + h.c. \notag \\
\end{eqnarray}
is the SSH model composed of $| j' \rangle$ and $| m \rangle$, $V_{j'}$ is the effective onsite energy of $| j' \rangle$ and $P_{j'} = \sum_{n}| n,j' \rangle \langle n,j' |$ is the projection operator over sublattice $| j' \rangle$. These $m-1$ SSH models aren't independent. As shown in Fig. \ref{fig:2}(a), they share the same sublattice $| m \rangle$. The coupling terms break the chiral symmetry of each SSH model:
\begin{eqnarray} \label{eq14}
	\Gamma^{\dagger}_{i'}\sum_{j'=1'}^{(m-1)'}H_{j'}^{SSH}\Gamma_{i'} \neq -\sum_{j'=1'}^{(m-1)'}H_{j'}^{SSH},
\end{eqnarray}
where $\Gamma_{i'} = \sum_{n}| n,i' \rangle \langle n,i' |- | n,m \rangle \langle n,m |$. 

However, the sub-symmetry \cite{wang2023sub} on the redefined sublattice $| j' \rangle$ is still satisfied:
\begin{eqnarray} \label{eq15}
	\Gamma^{\dagger}_{i'}\sum_{j'=1'}^{(m-1)'}H_{j'}^{SSH}\Gamma_{i'}P_{i'} = -\sum_{j'=1'}^{(m-1)'}H_{j'}^{SSH}P_{i'},
\end{eqnarray}
The sub-symmetry protects the zero-energy chiral edge states $| L_{i'} \rangle$ occupying sublattice $| i' \rangle$:
\begin{eqnarray} \label{eq75}
	\left\{\begin{matrix}
		\sum_{j'=1'}^{(m-1)'}H_{j'}^{SSH}| L_{i'} \rangle = 0
		\\\\
		P_{i'}| L_{i'} \rangle = | L_{i'} \rangle
	\end{matrix}\right..
\end{eqnarray}
Together with Eq. \eqref{eq11}, we can obtain:
\begin{eqnarray} \label{eq76}
	H_m'| L_{i'} \rangle = V_{i'}| L_{i'} \rangle.
\end{eqnarray}
Similar to the case of Rice-Mele model \cite{PhysRevB.103.205306,PhysRevB.110.104106}, the onsite terms $V_{i'}$ only shift the energy of the chiral edge state without affecting its wavefunction and topological property. Therefore, when $|v_{i'}| > |w_{i'}|$, i.e. $|{t_{m-1}\phi_{m-1}^{j'}}/{t_m\phi_1^{j'}}| < 1$, the sub-symmetry-protected chiral edge state with the energy of $V_{i'}$ can be found, which is consistent with the solution given by Eq. \ref{eq5}.

Following the same strategy, we can also illustrate the topological origin of the edge states at the right boundary $| R \rangle$. To be precise, here we perform another representation transformation to diagonalize $\mathcal{H}_{m-1}^{R}$ with the redefined sublattices $| j'' \rangle$. The Hamiltonian can then be written as:
\begin{eqnarray} \label{eq66}
	\mathcal{H}_{m}''(k) = \begin{bmatrix}
		0 & t_{1''}(k) & t_{2''}(k) & \cdots & t_{(m-1)''} \\
		t_{1''}^{\dagger}(k) & V_{1''} & & & \text{{\huge 0}} \\
		t_{2''}^{\dagger}(k) & & V_{2''} & & \\
		\vdots & & & \ddots & \\
		t_{(m-1)''}^{\dagger}(k) & \text{{\huge 0}} & & & V_{(m-1)''}
	\end{bmatrix}, \notag \\
\end{eqnarray}
where $t_{j''}(k)=w_{j''}+v_{j''}e^{ik}$ and $w_{j''}$ and $v_{j''}$ can be calculated by:
\begin{eqnarray} \label{eq67}
	\left\{\begin{matrix}
		w_{j''} = \langle n,1 | H_{m} | n,j'' \rangle
		\\\\
		v_{j''} = \langle n,j'' | H_{m} | n+1,1 \rangle 
	\end{matrix}\right..
\end{eqnarray}
As shown in Fig. \ref{fig:2}(b),$\mathcal{H}_{m}''(k)$ also contains $m-1$ coupled SSH models sharing sublattice $| 1 \rangle$. As a result, the chiral edge states occupying $| j'' \rangle$ are sub-symmetry-protected, which are exactly the right-boundary edge states given by Eq. \eqref{eq9}.

\subsection{\label{sec:level2}The Breaking of Sub-Symmetry by Inter-Cell Long-Range Hoppings in the Extended SSH\boldmath{$3$} Model}

\begin{figure}[t]
	\centering
	\includegraphics[width=1\linewidth]{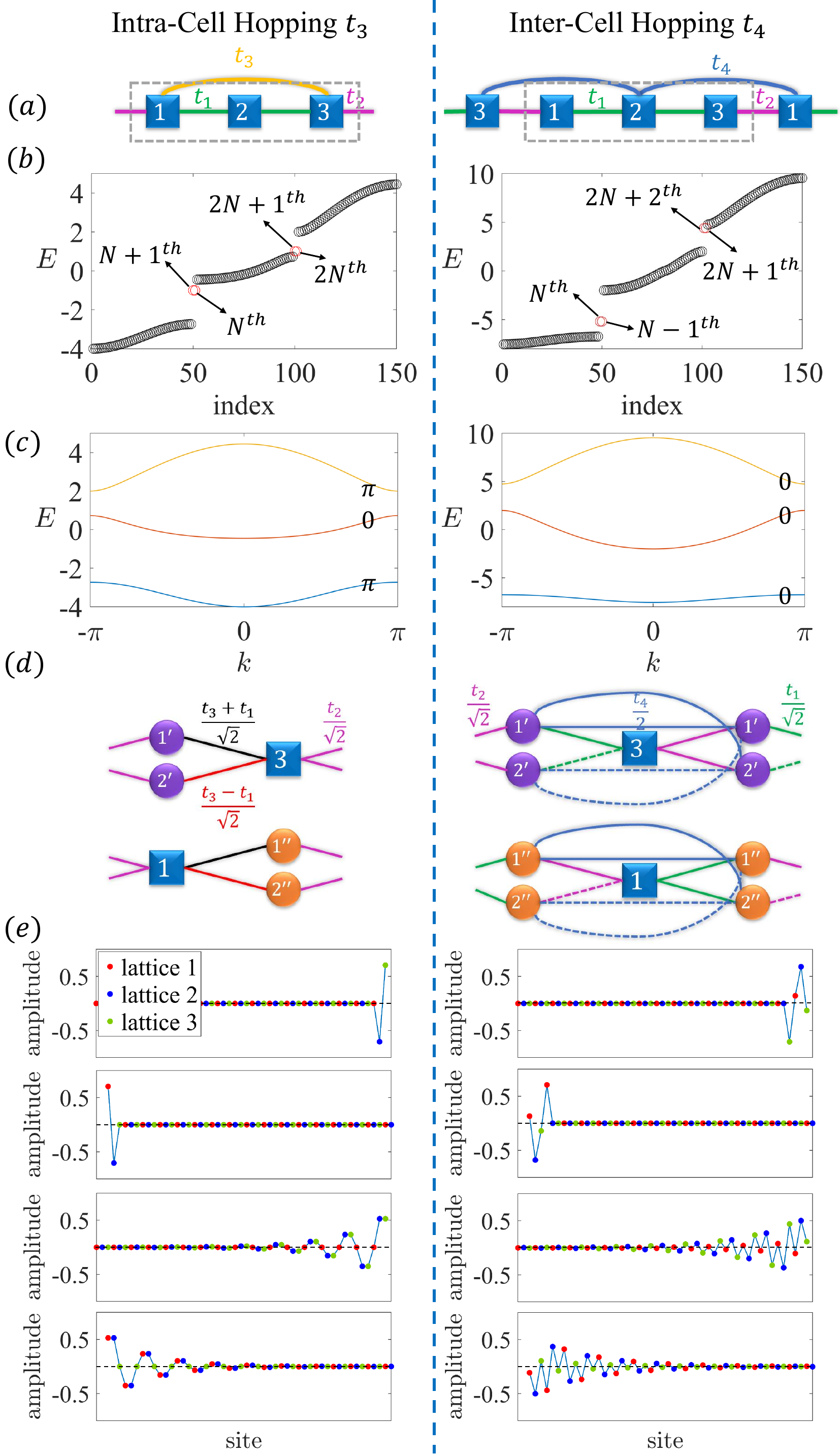}
	\caption{ (a) Schematics of the extended SSH$3$ models. In the left column, only the intra-cell hopping $t_3$ is considered. In the right column, only the inter-cell hopping $t_4$ is considered. In this section, the parameters are set to $t_1=1$, $t_2=3$, $t_3=1$ for the left column and $t_1=1$, $t_2=2$, $t_4=5$ for the right column. (b) The energy spectrum of the extended SSH$3$ models. The edge states are colored in red while their indexes are marked in the figure. (c) The bulk bands of the extended SSH$3$ models. The Zak phases are marked on the corresponding bands. (d) The schematics of the models with redefined sublattices. The dashed lines indicate negative hoppings. (e) The spatial distribution of the edge states. } \label{fig:12}
\end{figure}

While in this work we mainly focus on the systems with nearest-neighbour hoppings, we note that our method might also bring new insights to the models with long-range hoppings. In recent years, the topological property of the systems with long-range hoppings has attracted much attention \cite{PhysRevApplied.20.064042,PhysRevLett.131.157201,PhysRevLett.127.147401,PhysRevLett.128.127601,malakar2023engineering}, including the extended SSH$3$ model \cite{du2024one,PhysRevB.110.125424,ghuneim2024topological}. In previous study of the extended SSH$3$ model, the intra-cell and inter-cell long-range hoppings are equally treated while their differences are not highlighted. From the perspective of sub-symmetry protected topology and redefined sublattices, in this section we would like to distinguish these two kinds of long-range hoppings from the perspective of sub-symmetry protected topology.

As illustrated in Fig. \ref{fig:12}(a), here we consider the topological properties of the extended SSH$3$ model with only the intra-cell long-range hopping $t_3$ and the one with only the inter-cell long-range hopping $t_4$, respectively. The band structure under OBC is shown in Fig. \ref{fig:12}(b), where two pairs of two-fold degenerate edge states can be observed for both models.

However, we find that the topological properties of the edge states in these two cases are quite different. Since both of the extended SSH$3$ models considered here are mirror symmetric, we can analyze their topological properties from filling anomaly and the quantized Zak phases \cite{PhysRevB.108.085116}. In Fig. \ref{fig:12}(b), we find that the indexes of the edge states with $t_3$ are $N$, $N+1$, $2N$, $2N+1$, which satisfies the filling anomaly. In contrary, the indexes of the edge states with $t_4$ are $N-1$, $N$, $2N+1$, $2N+2$ and the filling anomaly is violated. In Fig. \ref{fig:12}(c), we also calculate the Zak phase for both cases, which further confirms that the edge states with $t_3$ are topologically nontrivial while the ones with $t_4$ are trivial.

The difference between intra-cell and inter-cell long-range hopping can be understood from the perspective of sub-symmetry. In Fig. \ref{fig:12}(d) we plot the schematics of both models with the redefined sublattices in Sec. \ref{II}(B). It clearly shows that the intra-cell long-range hopping $t_3$ only affects the value of effective hoppings while the sub-symmetry remains intact. The inter-cell long-range hopping $t_4$, on the other hand, introduces long-range hoppings in the redefined sublattices and therefore breaks the sub-symmetry. The difference between $t_3$ and $t_4$ also manifests itself in the spatial distribution of the edge states. As shown in Fig. \ref{fig:12}(e), the edge states of the extended SSH$3$ model with $t_3$ exhibit clear evidence of topological chiral edge states as the amplitude of the wavefunction on certain sublattice is pinned to zero. For the edge states with $t_4$, the chiral characteristics can not be observed.

It is worth noting that same conclusion can be drawn for any extended SSH$m$ model: As long as the long-range hoppings are kept intra-cell, the model can always be transformed into the form shown in \ref{fig:2}(a)-(b) while chiral edge states with zero amplitude at sublattice $1$ or $m$ can be observed. 

\section{\label{sec:level1}THE TOPOLOGICAL INVARIANT AND THE BULK-EDGE CORRESPONDENCE FOR 1D ASYMMETRIC SYSTEMS} \label{III}

In Sec. \ref{II}, we show the hidden sub-symmetry protected topological chiral edge states in the SSH$m$ model by redefining sublattices. However, as stated in Ref. \cite{wang2023sub}, defining a topological invariant in a system with sub-symmetry is not straightforward. In this section, based on isospectral reduction, we propose a proper topological invariant to establish the bulk-edge correspondence. For the asymmetric model with chiral edge states, we show that its effective Hamiltonian might have a Rice-Mele-like form. Then, the mirror symmetry of its reduced wavefuntion can be reconstructed through normalization. Therefore, a quantized $\bar{Z}$ can be calculated by the normalized reduced wavefunction. Our new topological invariant can be utilized in more complex systems beyond the simplest 1D chain models. As an example, we introduce our topological invariant to 1D asymmetric ladder SSH model, and the bulk-edge correspondence can be established.

\subsection{\label{sec:level2}Chiral Edge States in Rice-Mele-Like Effective Hamiltonian}

To begin with, we would like to briefly introduce the concept of isospectral reduction proposed in latent symmetry \cite{PhysRevLett.126.180601,PhysRevB.110.035106,10.21468/SciPostPhys.18.2.061}. For a system with a Hamiltonian $\mathcal{H}(k)$, the sublattices in each unit cell can be divided into two sets: $S$ and its complement $\bar{S}$. Here set $S$ is composed of all the sublattices related to the inter-cell hopping terms. Then the Hamiltonian can be represented as:
\begin{eqnarray} \label{eq22}
	\mathcal{H}(k) = \begin{bmatrix}
		\mathcal{H}_{SS}(k) & \mathcal{H}_{S\bar{S}}\\
		\mathcal{H}_{\bar{S}S} & \mathcal{H}_{\bar{S}\bar{S}}
	\end{bmatrix}.
\end{eqnarray}
The eigenequation of $\mathcal{H}(k)$ can be reduced into the generalized eigenequation of set $S$:
\begin{eqnarray} \label{eq23}
	\tilde{\mathcal{H}}_S(k,E)\phi_S(k) = E\phi_S(k)
\end{eqnarray}
with
\begin{eqnarray} \label{eq24}
	\tilde{\mathcal{H}}_S(k,E) = \mathcal{H}_{SS} - \mathcal{H}_{S\bar{S}}(\mathcal{H}_{\bar{S}\bar{S}}-EI)^{-1}\mathcal{H}_{\bar{S}S}.
\end{eqnarray}
If $\tilde{\mathcal{H}}_S(k,E)$ possesses mirror symmetry, $\mathcal{H}(k)$ is referred to as being latent symmetric \cite{PhysRevLett.126.180601}.

\begin{figure}[t]
	\centering
	\includegraphics[width=1\linewidth]{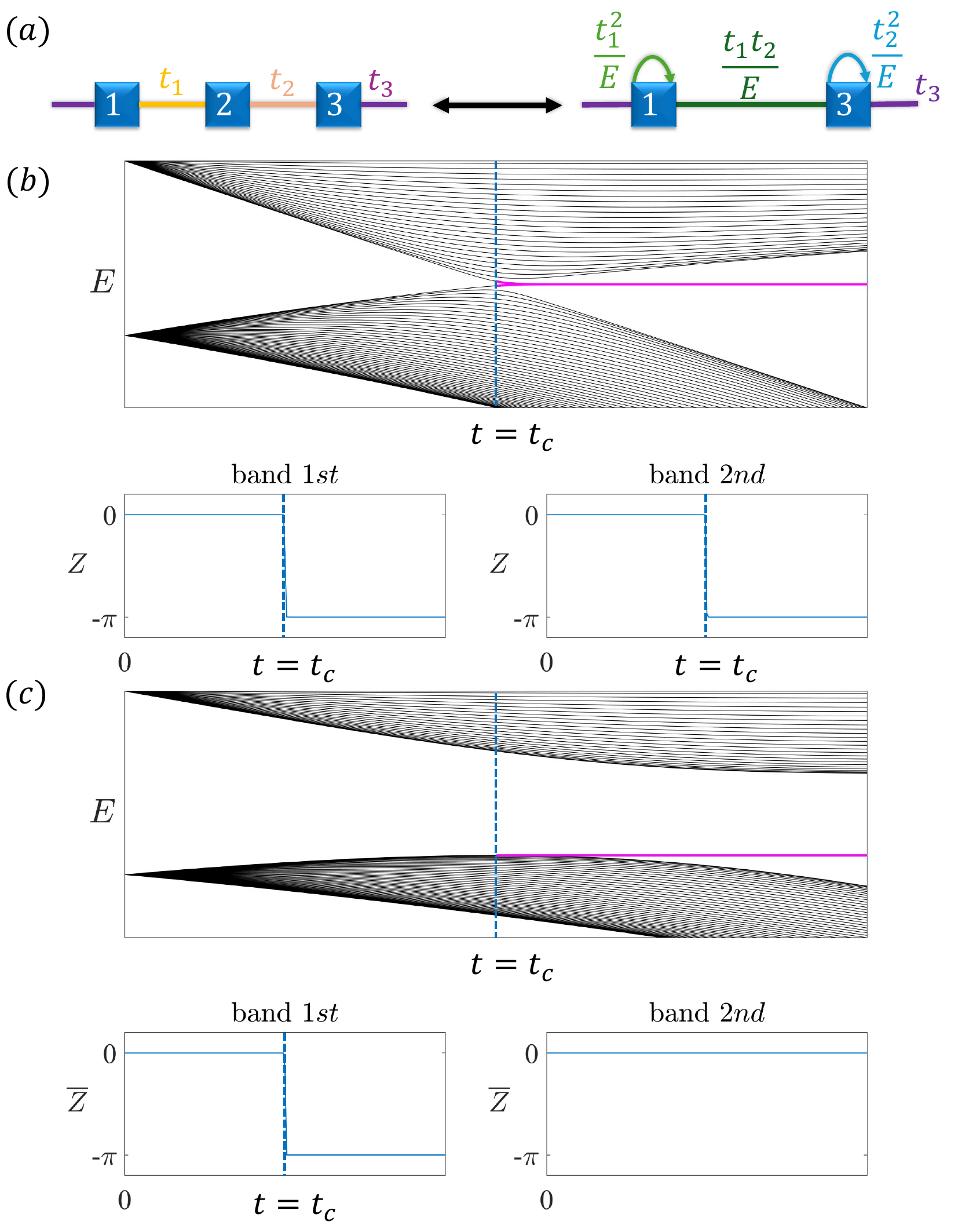}
	\caption{ (a) Schematic of the effective Hamiltonian of SSH$3$ model with $S = \{| 1 \rangle, | 3 \rangle \}$. (b)-(c) Schematic of the topological phase transition (b) with gap closing and (c) without gap closing. Here $t$ stands for the systems parameter where $t_c$ is the critical point. The corresponding topological edge states are marked by magenta lines. The topological invariant, i.e. the conventional Zak phase $Z$ and the normalized Zak phase $\bar{Z}$ are also plotted in (b) and (c), respectively. } \label{fig:4}
\end{figure}

The merit of isospectral reduction is to describe a given system with minimum sublattices while the intra-cell structure is neglected. Comparing with the original Hamiltonian, the effective Hamiltonian $\tilde{\mathcal{H}}_S$ might reveal the properties of the system in a more fundamental way. For the SSH$m$ model discussed in Sec. \ref{II}, we can obtain $\tilde{\mathcal{H}}_S$ with $S = \{| 1 \rangle, | m \rangle \}$. In this case $\tilde{\mathcal{H}}_S(k)$ takes the form of Rice-Mele model:
\begin{eqnarray} \label{eq68}
	\tilde{\mathcal{H}}_S(k,E) = \begin{bmatrix}
		v_1(E) & h(k,E)\\
		h^{\dagger}(k,E) & v_2(E)
	\end{bmatrix},
\end{eqnarray}
which supports chiral edge states when $E=v_1(E)$ or $E=v_2(E)$. These chiral edge states have zero amplitude at sublattice $m$ or $1$, which exactly corresponds to the $| L \rangle$ or $| R \rangle$ obtained in Sec. \ref{II}(A), respectively. Taking the SSH$3$ model as an example, we have $S = \{| 1 \rangle, | 3 \rangle \}$. As shown in Fig. \ref{fig:4}(a), $\tilde{\mathcal{H}}_S(k,E)$ reads as \cite{PhysRevApplied.23.L031001,10.1063/5.0256334}:
\begin{eqnarray} \label{eq25}
	\tilde{\mathcal{H}}_S(k,E) = \begin{bmatrix}
		\frac{t_1^2}{E} & \frac{t_1t_2}{E}+t_3e^{ik}\\
		\frac{t_1t_2}{E}+t_3e^{-ik} & \frac{t_2^2}{E}
	\end{bmatrix},
\end{eqnarray}
where chiral edge states can be found at $\frac{t_1^2}{E}=E$ or $\frac{t_2^2}{E}=E$, i.e. $E=\pm t_1$ or $E=\pm t_2$, which are exactly the energies of the edge states shown in Fig. \ref{fig:1}(b).

While we have shown that the chiral edge states in SSH$m$ model originates from its Rice-Mele effective Hamiltonian, for the more complicated 1D systems (e.g. ladder model) or even 2D systems, chiral edge states can also be found if its effective Hamiltonian take a Rice-Mele-like form:
\begin{eqnarray} \label{eq26}
	\tilde{\mathcal{H}}_S(k,E) = \begin{bmatrix}
		v_1(E)I & h(k,E)\\
		h^{\dagger}(k,E) & v_2(E)I
	\end{bmatrix},
\end{eqnarray}
here $h(k,E)$ can be a matrix of any dimension, $v_1(E)$, $v_2(E)$ are the effective onsite terms, and $I$ is the identity matrix. Its generalized eigenequation reads as:
\begin{eqnarray} \label{eq27}
	\begin{bmatrix}
		v_1(E)I & h(k,E)\\
		h^{\dagger}(k,E) & v_2(E)I
	\end{bmatrix}
	\begin{bmatrix}
		\phi_A(k)\\
		\phi_B(k)
	\end{bmatrix} = E \begin{bmatrix}
		\phi_A(k)\\
		\phi_B(k)
	\end{bmatrix},
\end{eqnarray}
where the sublattice set $S$ is further divided into two sets $A$ and $B$ ($A = \{| 1 \rangle\}$ and $B= \{| m \rangle\}$ for SSH$m$ model). In this case, $\tilde{\mathcal{H}}_S(k,E)$ also supports two chiral edge states when $E=v_1(E)$ and $E=v_2(E)$. These chiral edge states have zero amplitude over sublattice $A$ and sublattice $B$ respectively, which are exactly the chiral edge states of the original system.

Through the effective Hamiltonian, the difference between intra-cell and inter-cell long-range hoppings in the extended SSH$m$ model can also be understood. Based on the definition of set $S$, intra-cell long-range hoppings keep $S$ unchanged. Therefore, $\tilde{\mathcal{H}}_S$ is still in Rice-Mele form and supports chiral edge states. In contrary, the existence of inter-cell long-range hoppings will introduce additional sublattices into $S$, which consequently breaks the Rice-Mele form of $\tilde{\mathcal{H}}_S$.

\subsection{\label{sec:level2}The Normalized Zak Phase \boldmath{$\bar{Z}$}}

For a given system with Rice-Mele-like effective Hamiltonian $\tilde{\mathcal{H}}_S(k,E)$, we define $\tilde{\mathcal{H}}_S^0(k,E)$ as its counterpart without onsite terms:
\begin{eqnarray} \label{eq70}
	\tilde{\mathcal{H}}_S^0(k,E) = \begin{bmatrix}
	0 & h(k,E)\\
	h^{\dagger}(k,E) & 0
	\end{bmatrix}.
\end{eqnarray}
If $\tilde{\mathcal{H}}_S^0(k,E)$ further possesses mirror symmetry, i.e.:
\begin{eqnarray} \label{eq28}
	M_x^{-1} \tilde{\mathcal{H}}_S^0(k,E) M_x = \tilde{\mathcal{H}}_S^0(-k,E),
\end{eqnarray}
where $M_x = \sigma_x \otimes I$, we can define the normalized reduced wavefunction over set $A$ and $B$ respectively:
\begin{eqnarray} \label{eq29}
	\tilde{\psi}_S(k) = \begin{bmatrix}
		\tilde{\phi}_A(k)\\
		\tilde{\phi}_B(k)
	\end{bmatrix}, \quad \tilde{\phi}_\sigma(k)=\frac{1}{\sqrt{2\langle \phi_\sigma | \phi_\sigma \rangle}}\phi_\sigma(k),
\end{eqnarray}
with $\sigma = A,B$. It can be proved that $\tilde{\psi}_S(k)$ is mirror symmetric, i.e. the normalized reduced wavefunction at $\pm k$ is related by $M_x$ up to a phase factor $\theta(k)$ [Appendix \ref{A}]:
\begin{eqnarray} \label{eq30}
	M_x \tilde{\psi}_S(k) = e^{i\theta(k)}\tilde{\psi}_S(-k)
\end{eqnarray}
Therefore, the Zak phase defined by $\tilde{\psi}_S(k)$ is quantized \cite{PhysRevLett.126.113901}:
\begin{eqnarray} \label{eq31}
	\bar{Z}_j=i\int dk \langle \tilde{\psi}_S^j(k) | \partial_k \tilde{\psi}_S^j(k) \rangle = 0 \ or \ \pi,
\end{eqnarray}
where $j$ is the band index.

For the conventional SPT phases, as illustrated in Fig. \ref{fig:4}(b), the topological phase transition takes place at the gap closing point. As a result, the degenerate edge states always come in pairs and the Zak phases of adjacent bands always change simultaneously. With the normalized Zak phase $\bar{Z}$, we find another type of topological phase transition in asymmetric systems. As shown in Fig. \ref{fig:4}(c), accompanied by the appearance of chiral edge states, the
normalized Zak phase $\bar{Z}$ of the jth band takes an abrupt change. Compared with the conventional SPT phase transition, the normalized Zak phase of each
band can change independently and the edge states do not have to come in pairs. These features are in good match with the topological properties of chiral edge states in asymmetric systems, indicating the bulk-edge correspondence can be established.

We note that $\bar{Z}$ is also a $\mathbb{Z}_2$ invariant. That is to say, $\bar{Z}$ remains unchanged when even number of edge states emerge simultaneously from the same band (e.g. the second band of SSH$3$ model).

\subsection{\label{sec:level2}The Bulk-Edge Correspondence in Asymmetric Ladder SSH Model}

\begin{figure}[t]
	\centering
	\includegraphics[width=1\linewidth]{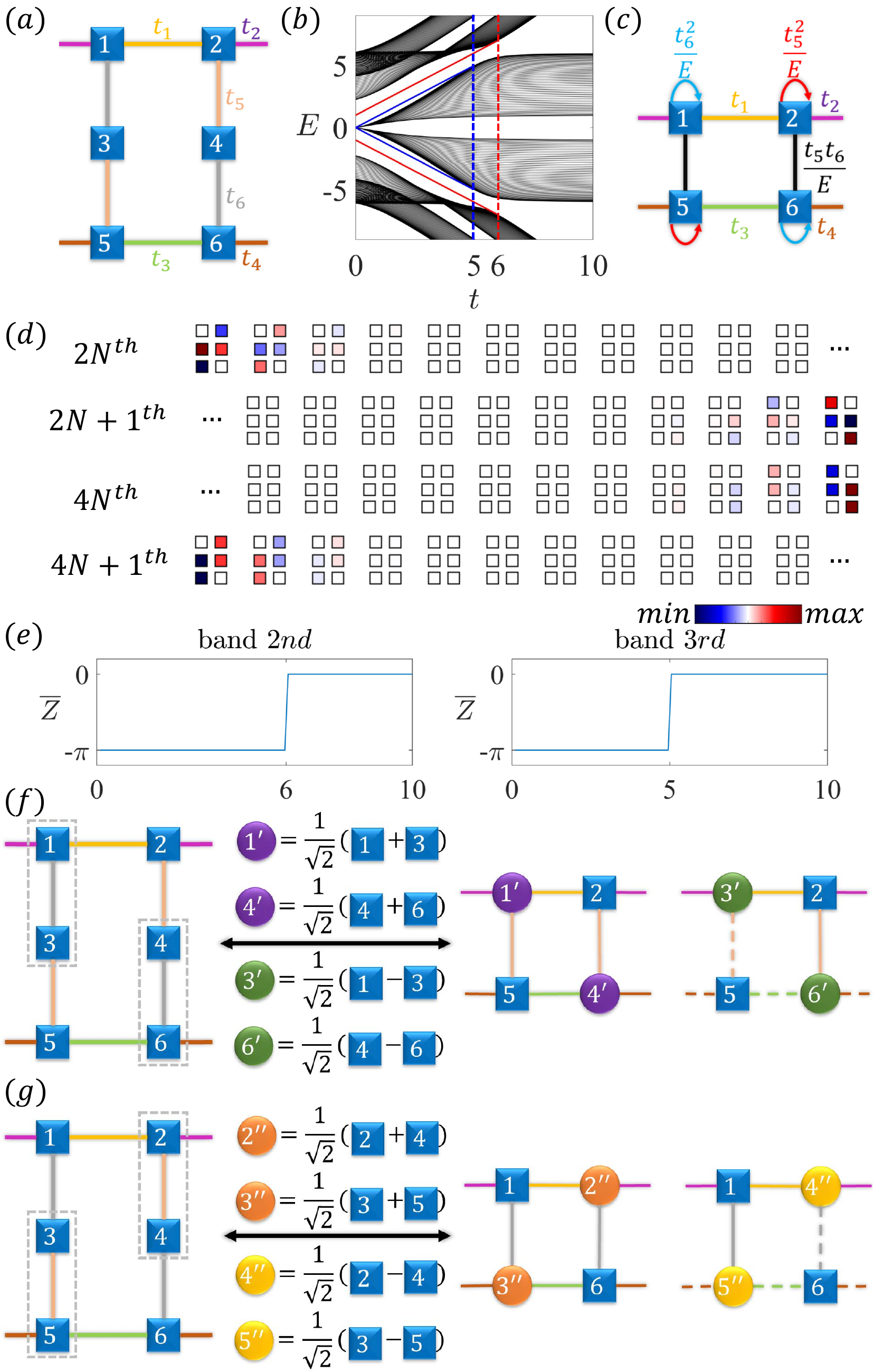}
	\caption{ (a) The schematic of the asymmetric ladder SSH model. (b) The band structure versus $t$ under OBC with $t_1=5$, $t_2=1$, $t_3=2$, $t_4=4$, $t_5=t+1$, $t_6=t$. The edge states are marked by blue and red lines. (c) The schematic of the effective Hamiltonian with $S=\{| 1 \rangle, | 2 \rangle, | 5 \rangle, | 6 \rangle \}$. (d) The spatial distribution of the edge states of the asymmetric ladder SSH model, showing typical characteristics of chiral edge states. (e)-(f) Schematic of the redfined sublattices diagonalizing the subspace composed of (e) $\{| 1 \rangle, | 3 \rangle, | 4 \rangle, | 6 \rangle \}$; (f) $\{| 2 \rangle, | 3 \rangle, | 4 \rangle, | 5 \rangle \}$. In the redefined sublattices, the onsite energies are distinguished by different colors while the hopping terms are $1/\sqrt{2}$ of the ones sharing the same color in the original sublattices. The dashed lines indicate negative hoppings. (g) The normalized Zak phase $\bar{Z}$ of the second and third band. } \label{fig:5}
\end{figure}

The advantage of our new invariant $\bar{Z}$ is that it can be utilized in more complicated systems, e.g. the 1D ladder models or even 2D models. As an example, in this section we would like to extend $\bar{Z}$ into the asymmetric ladder SSH model and establish the bulk-edge correspondence. As shown in Fig. \ref{fig:5}(a), the Hamiltonian of asymmetric ladder SSH model can be written as:
\begin{eqnarray} \label{eq50}
	\mathcal{H}_{ladder}(k) = \begin{bmatrix}
		0 & l_1 & t_6 & 0 & 0 & 0 \\
		l_1^* & 0 & 0 & t_5 & 0 & 0 \\
		t_6 & 0 & 0 & 0 & t_5 & 0 \\
		0 & t_5 & 0 & 0 & 0 & t_6 \\
		0 & 0 & t_5 & 0 & 0 & l_2 \\
		0 & 0 & 0 & t_6 & l_2^* & 0
	\end{bmatrix},
\end{eqnarray}
where $l_1=t_1+t_2e^{ik}$, $l_2=t_3+t_4e^{ik}$. When $t_5 \neq t_6$, the spatial inversion symmetry is broken; When $t_1 \neq t_3$ and $t_2 \neq t_4$, the $C_2$ rotation symmetry is broken. The breaking of spatial symmetry makes the conventional Zak phase unquantized. On the other hand, $\mathcal{H}_{ladder}(k)$ possesses chiral symmetry, which makes the band structure symmetric around $E=0$.

The band structure versus $t$ under OBC with $t_1=5$, $t_2=1$, $t_3=2$, $t_4=4$, $t_5=t+1$, $t_6=t$ is shown in Fig. \ref{fig:5}(b). Two pairs of edge states marked by red and blue lines can be observed in the second and fourth gaps. The energies of the edge states are $E=\pm t_5$ and $E=\pm t_6$ while the critical points are $t=5$ and $t=6$, respectively. The spatial distribution of the edge states are shown in Fig. \ref{fig:5}(d), which exhibits characteristics of chiral edge states in a way that only part of the sublattices are occupied.

We first consider the effective Hamiltonian $\tilde{\mathcal{H}}_S(k,E)$ with $S=\{| 1 \rangle, | 2 \rangle, | 5 \rangle, | 6 \rangle \}$. As shown in Fig. \ref{fig:5}(c), in the base $A = \{| 1 \rangle, | 6 \rangle \}$ and $B = \{| 2 \rangle, | 5 \rangle \}$, $\tilde{\mathcal{H}}_S(k,E)$ exhibits a Rice-Mele form:
\begin{eqnarray} \label{eq51}
	\tilde{\mathcal{H}}_S(k,E) = \begin{bmatrix}
		\frac{t_6^2}{E}I_2 & h(k,E)\\
		h^{\dagger}(k,E) & \frac{t_5^2}{E}I_2
	\end{bmatrix},
\end{eqnarray}
with
\begin{eqnarray} \label{eq52}
	h(k,E) = \begin{bmatrix}
		t_1+t_2e^{ik} & \frac{t_5t_6}{E}\\
		\frac{t_5t_6}{E} & t_3+t_4e^{-ik}
	\end{bmatrix}.
\end{eqnarray}
When $E=\frac{t_6^2}{E}$ or $E=\frac{t_5^2}{E}$, the system supports chiral edge states with zero amplitude over set $B$ or $A$ respectively, which are exactly the edge states in Fig. \ref{fig:5}(d). 

As to the topological invariant, since the effective Hamiltonian without onsite terms $\tilde{\mathcal{H}}_S^0(k)$ is mirror symmetric, $\bar{Z}$ can be calculated by the normalized reduced wavefunction $\tilde{\psi}_S(k)$. In Fig. \ref{fig:5}(e), $\bar{Z}$ of the second (third) band exhibits an abrupt change at $t=6$ ($t=5$), which coincides with the appearance of the chiral edge states shown in Fig. \ref{fig:5}(b).

The topological origin of the edge states can be further understood through the sub-symmetry protected topology with redefined sublattices. For the edge states with zero amplitude over set $B = \{| 2 \rangle, | 5 \rangle \}$, we perform the representation transformation to diagonalize the subspace composed of the complement set $\bar{B} = \{| 1 \rangle, | 3 \rangle , | 4 \rangle, | 6 \rangle \}$:
\begin{eqnarray} \label{eq72}
	\left\{\begin{matrix}
		| 1' \rangle = \frac{1}{\sqrt{2}}(| 1 \rangle + | 3 \rangle), \ V_{1'} = t_6
		\\\\
		| 4' \rangle = \frac{1}{\sqrt{2}}(| 4 \rangle + | 6 \rangle), \ V_{4'} = t_6
		\\\\
		| 3' \rangle = \frac{1}{\sqrt{2}}(| 1 \rangle - | 3 \rangle), \ V_{3'} = -t_6
		\\\\
		| 6' \rangle = \frac{1}{\sqrt{2}}(| 4 \rangle - | 6 \rangle), \ V_{6'} = -t_6
	\end{matrix}\right..
\end{eqnarray}
As shown in Fig. \ref{fig:5}(f), the Hamiltonian after transformation can be written as:
\begin{eqnarray} \label{eq53}
	H_{ladder}' = H^{RM}_{\{1',2,5,4'\}} + H^{RM}_{\{3',2,5,6'\}}
\end{eqnarray}
where $H^{RM}_{\{\alpha,\beta,\gamma,\delta\}}$ stands for the Rice-Mele ladder \cite{PhysRevB.110.104106} defined on sublattices $| \alpha \rangle$, $| \beta \rangle$, $| \gamma \rangle$, $| \delta \rangle$. It is known that Rice-Mele ladder supports chiral edge states occupying the opposite sublattices. Since these two Rice-Mele ladders share the same sublattices $| 2 \rangle$ and $| 5 \rangle$, based on the analysis in Sec. \ref{II}(B), the chiral edge states occupying sublattice $| 1' \rangle$, $| 4' \rangle$ and $| 3' \rangle$, $| 6' \rangle$ are sub-symmetry protected. The detailed derivation can be found in Appendix \ref{E}. These are exactly the $4N^{th}$ and ${(2N+1)}^{th}$ states shown in Fig. \ref{fig:5}(d). 

Similarly, for the edge states with zero amplitude over set $A$, we diagonalize the subspace composed of the complement set $\bar{A} = \{| 2 \rangle, | 3 \rangle , | 4 \rangle, | 5 \rangle \}$:
\begin{eqnarray} \label{eq73}
	\left\{\begin{matrix}
		| 2'' \rangle = \frac{1}{\sqrt{2}}(| 2 \rangle + | 4 \rangle), \ V_{2''} = t_5
		\\\\
		| 3'' \rangle = \frac{1}{\sqrt{2}}(| 3 \rangle + | 5 \rangle), \ V_{3''} = t_5
		\\\\
		| 4'' \rangle = \frac{1}{\sqrt{2}}(| 2 \rangle - | 4 \rangle), \ V_{4''} = -t_5
		\\\\
		| 5'' \rangle = \frac{1}{\sqrt{2}}(| 3 \rangle - | 5 \rangle), \ V_{5''} = -t_5
	\end{matrix}\right..
\end{eqnarray}
which also results in two coupled Rice-Mele ladders sharing sublattices $| 1 \rangle$, $| 6 \rangle$ shown in Fig. \ref{fig:5}(g):
\begin{eqnarray} \label{eq74}
	H_{ladder}'' = H^{RM}_{\{1,2'',3'',6\}} + H^{RM}_{\{1,4'',5'',6\}}.
\end{eqnarray}
In this case, the chiral edge states occupying sublattice $| 2'' \rangle$, $| 3'' \rangle$ and $| 4'' \rangle$, $| 5'' \rangle$ are sub-symmetry protected, which correspond to the ${(4N+1)}^{th}$ and $2N^{th}$ states in Fig. \ref{fig:5}(d).

\section{\label{sec:level1}TOPOLOGICAL CHIRAL CORNER STATES WITH REDEFINED SUBLATTICES IN 2D SYSTEMS} \label{IV}

In this section, we would like to extend our theory into 2D systems where a new kind of higher-order topology can be found. With redefined sublattices, new topological chiral corner states protected by sub-symmetry are discovered and new topological phases can be defined. These corner states remain intact even when the spatial symmetry is fully broken, exhibiting a fundamental difference with the conventional quadrupole corner states. Here we construct the 2D BBH$3$ model as a prototype to exhibit our findings. By breaking the spatial symmetry, higher-order topological phase transition without gap closing is observed. Even more, some of the chiral corner states can move into the bulk continuum, exhibiting typical characteristics of the TBICs. As to the topological invariant, we introduce our new topological invariant $\bar{Z}$ into 2D and the edge-corner correspondence is established through the normalized edge polarization $\bar{p}^{edge}$.

\subsection{\label{sec:level2}BBH3 Model}

\begin{figure}[t]
	\centering
	\includegraphics[width=1\linewidth]{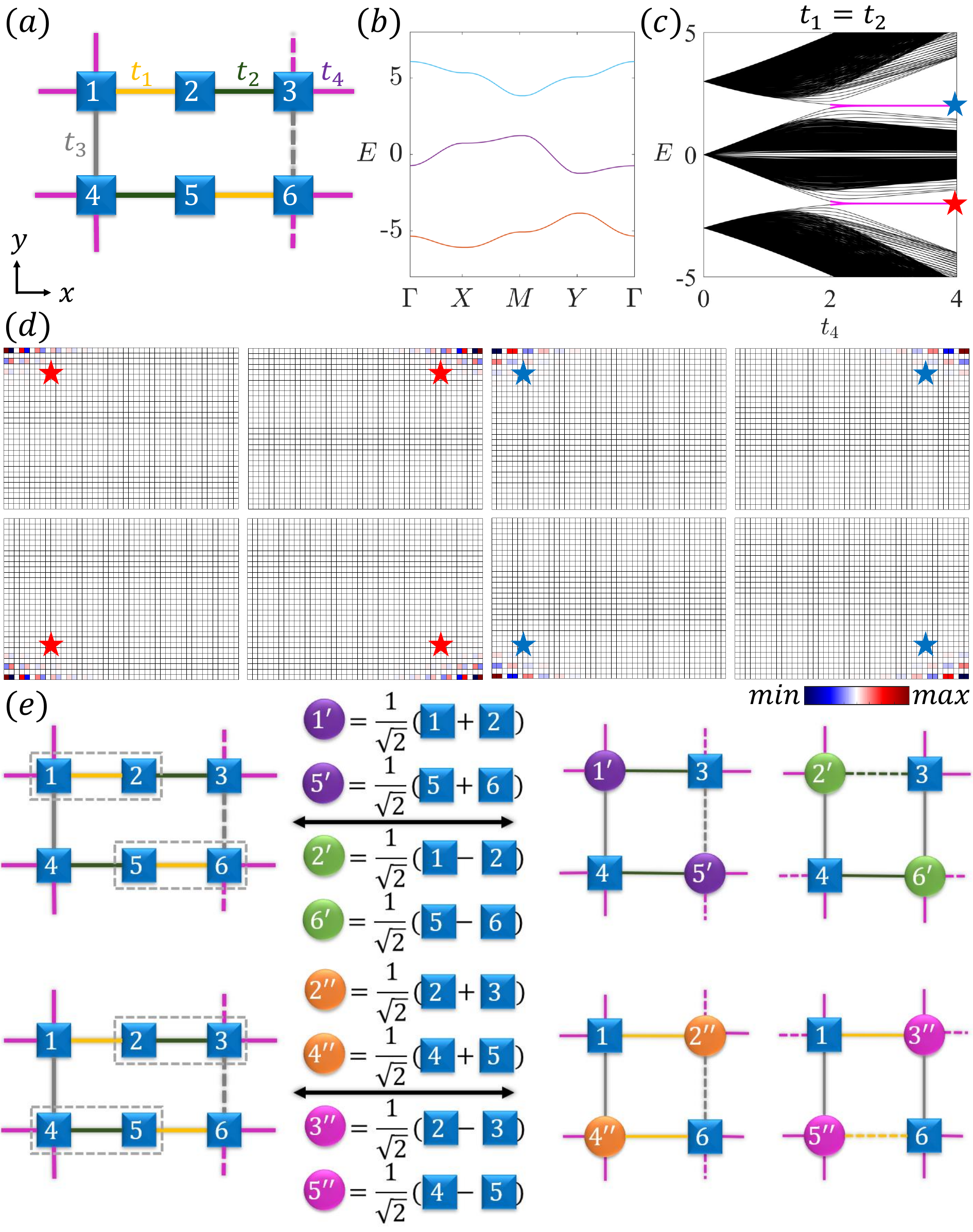}
	\caption{ (a) Schematic of the BBH$3$ model. The dashed lines indicate negative hoppings. (b) The bulk bands of the BBH$3$ model. All the bands are two-fold degenerate over the whole Brillouin Zone. (c) The band structure versus $t_4$ under OBC with $t_1=t_2=2$, $t_3=1$. (d) The spatial distribution of the corner states, exhibiting typical characteristics of chiral corner states. (e) Schematics of the BBH$3$ model with two types of redefined sublattices. } \label{fig:6}
\end{figure}

Here we consider a 2D model whose unit-cell structure is shown in Fig. \ref{fig:6}(a). It can be regarded as stacking the SSH$3$ model and its mirror-symmetric counterpart alternatively along $y$ direction. Mimic the BBH model \cite{doi:10.1126/science.aah6442}, here we introduce a $\pi$ flux in each plaquette by setting negative hoppings in $y$ direction, which are marked by dashed lines in Fig. \ref{fig:6}(a). In this sense, we define this model as the BBH3 model. The Hamiltonian of BBH$3$ model can be written as:
\begin{eqnarray} \label{eq32}
	\mathcal{H}_{BBH3}(k_x,k_y) = \begin{bmatrix}
		0 & t_1 & h_1 & h_2 & 0 & 0 \\
		t_1 & 0 & t_2 & 0 & 0 & 0 \\
		h_1^* & t_2 & 0 & 0 & 0 & -h_2 \\
		h_2^* & 0 & 0 & 0 & t_2 & h_1 \\
		0 & 0 & 0 & t_2 & 0 & t_1 \\
		0 & 0 & -h_2^* & h_1^* & t_1 & 0
	\end{bmatrix},
\end{eqnarray}
where $h_1 = t_4e^{ik_x}$, $h_2 = t_3+t_4e^{-ik_y}$. The bulk bands of the BBH$3$ model are shown in Fig. \ref{fig:6}(b).

In terms of symmetry, BBH$3$ model has projective inversion symmetry $P_I^{-1}\mathcal{H}_{BBH3}(k_x,k_y)P_I = \mathcal{H}_{BBH3}(-k_x,-k_y)$ with  $P_I^2 = -1$. Along with the time-reversal symmetry $T$, it gives rise to two-fold band Kramers degeneracy over the whole Brillouin Zone \cite{PhysRevLett.132.213801}. When $t_1=t_2$, BBH$3$ model is also mirror symmetric in both $x$ and $y$ directions.

In the following, we would like to discuss the higher-order topological properties of the BBH$3$ model under three different settings. In Sec. \ref{IV}(B)-(C), we first consider the mirror-symmetric case, i.e $t_1=t_2$. In Sec. \ref{IV}(D), we investigate the general situation where $t_1 \neq t_2$. Additionally, in Sec. \ref{IV}(E), we introduce the asymmetric BBH$3$ model where all the spatial symmetry is broken.

\subsection{\label{sec:level2}Chiral Corner States in Mirror-Symmetric BBH3 Model}

In this section, we first consider the corner states in mirror-symmetric BBH$3$ model by setting $t_1=t_2$. The band structure versus $t_4$ under OBC with $t_1=t_2=2$, $t_3=1$ is shown in Fig. \ref{fig:6}(c). When $t_4$ passes the critical point, accompanied by the closing and reopening of both gaps, four-fold degenerate corner states localized at four corners can be observed in each gap, which are marked by magenta lines in Fig. \ref{fig:6}(c). The indexes of the eight corner states are $2N^2-1$, $2N^2$, $2N^2+1$, $2N^2+2$ and $4N^2-1$, $4N^2$, $4N^2+1$, $4N^2+2$ respectively, indicating a filling anomaly \cite{PhysRevB.108.085116} which further confirms their nontrivial topology. In Fig. \ref{fig:6}(d), we exhibit the spatial distribution of these corner states. All of them show typical chiral characteristics in a way that the amplitudes of the wavefunctions at certain sublattices remain zero.

To investigate the topological origins of these corner states, we redefine the sublattices in a way similar to Sec. \ref{III}(C). As shown in Fig. \ref{fig:6}(e), we first take the representation transformation to diagonalize the subspace composed of $\{ | 1 \rangle, | 2 \rangle, | 5 \rangle, | 6 \rangle \}$, which results in two coupled 2D Rice-Mele models \cite{PhysRevB.110.104106} sharing sublattices $| 3 \rangle$ and $| 4 \rangle$:
\begin{eqnarray} \label{eq54}
	H_{BBH3}' = H^{2DRM}_{\{1',3,4,5'\}} + H^{2DRM}_{\{2',3,4,6'\}}
\end{eqnarray}
where $H^{2DRM}_{\{\alpha,\beta,\gamma,\delta\}}$ stands for the 2D Rice-Mele model defined on sublattices $| \alpha \rangle$, $| \beta \rangle$, $| \gamma \rangle$, $| \delta \rangle$. It is known that 2D Rice-Mele model supports four chiral corner states occupying one of the sublattices. Therefore, the four chiral corner states occupying the redefined sublattices $| 1' \rangle$, $| 2' \rangle$, $| 5' \rangle$, $| 6' \rangle$ respectively are sub-symmetry protected. 

As to the other four corner states, we perform another representation transformation to diagonalize the subspace composed of $\{ | 2 \rangle, | 3 \rangle, | 4 \rangle, | 5 \rangle \}$. The Hamiltonian after transformation reads as:
\begin{eqnarray} \label{eq55}
	H_{BBH3}'' = H^{2DRM}_{\{1,2'',4'',6\}} + H^{2DRM}_{\{1,3'',5'',6\}}
\end{eqnarray}
and the four chiral corner states occupying the redefined sublattices $| 2'' \rangle$, $| 4'' \rangle$, $| 5'' \rangle$, $| 6'' \rangle$ respectively are also sub-symmetry protected. Therefore, all of the eight corner states shown in Fig. \ref{fig:6}(c) and Fig. \ref{fig:6}(d) are chiral corner states with redefined sublattices.

\subsection{\label{sec:level2}The Edge-Corner Correspondence for Mirror-Symmetric BBH3 Model}

\begin{figure}[t]
	\centering
	\includegraphics[width=1\linewidth]{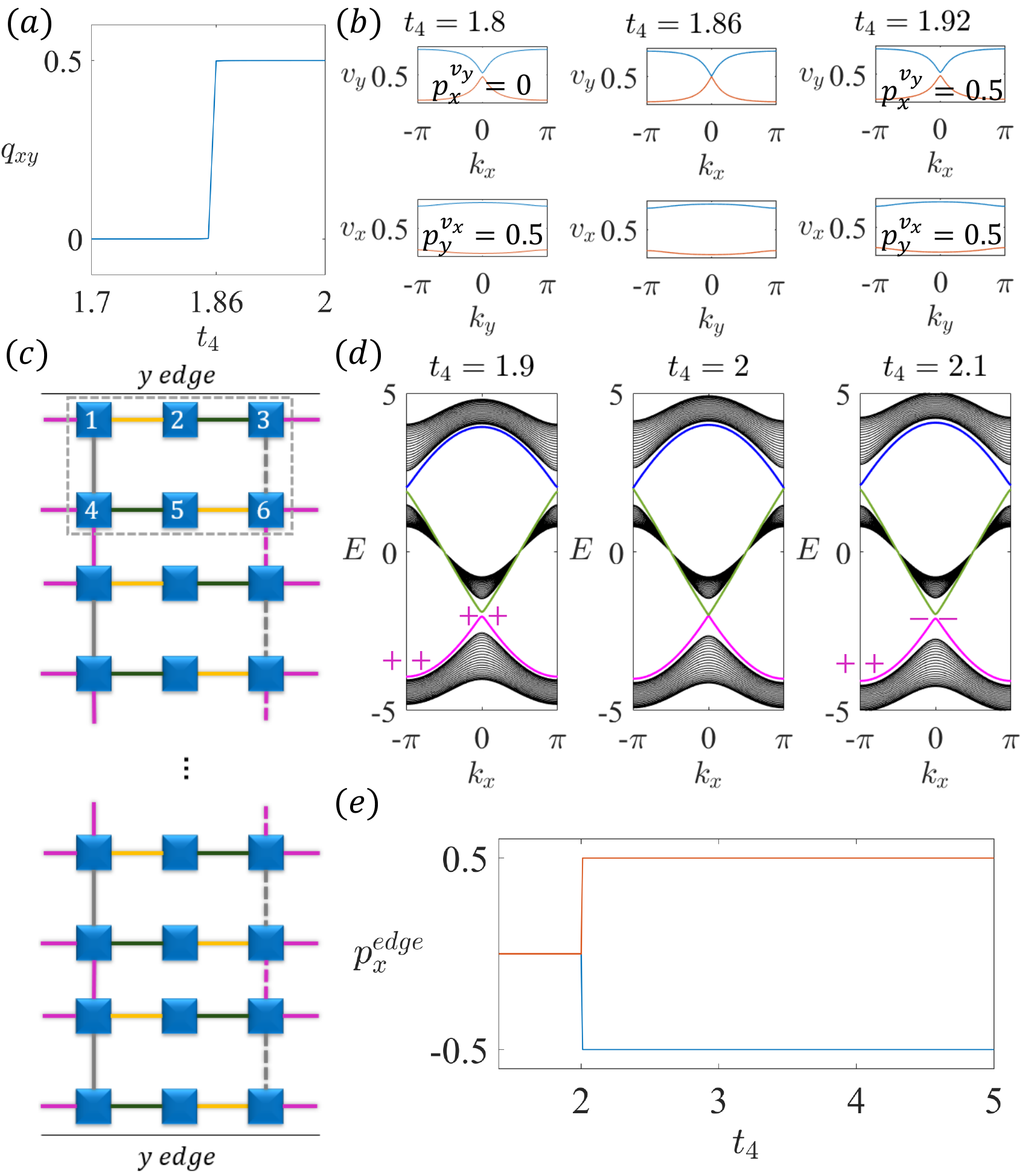}
	\caption{ The bulk-corner and edge-corner correspondence of the mirror-symmetric BBH$3$ model. (a) The quadrupole moment $q_{xy}$ versus $t_4$ at $1/3$ filling. (b) The Wannier band $v_x(k_y)$ and $v_y(k_x)$ at $1/3$ filling, showing a topological phase transition at $t_4=1.86$. (c) Schematic of the $y$-open ribbon structure of the BBH$3$ model. (d) The band structure of the $y$-open system in (c). Three pairs of two-fold degenerate edge bands are plotted in different colors while the parities of the lowest two edge bands at high-symmetry points are marked by $\pm$ signs. (e) The edge polarization of the lowest two edge bands with the $y$-open ribbon structure versus $t_4$. } \label{fig:7}
\end{figure}

Generally speaking, the higher-order topology can be described by two methods. One is the bulk-corner correspondence where the topological property of the corner state is illustrated by the quantized bulk invariant, such as the 2D Zak phase \cite{PhysRevLett.118.076803} and the quadrupole moment \cite{PhysRevB.96.245115}. The other is the edge-corner correspondence which relies on the topological properties of the edge bands under ribbon structure, such as the edge polarization \cite{PhysRevResearch.2.033029,PhysRevLett.132.213801} and the effective boundary Hamiltonian \cite{PhysRevB.102.121405}. In this section, we surprisingly find that the bulk-corner correspondence fails in the mirror-symmetric BBH$3$ model even though a quantized bulk invariant can be calculated. However, the edge-corner correspondence can still be established by the edge polarization.

Firstly, we would like to exhibit the failure of bulk topological invariant. The presence of mirror symmetry $M_x$, $M_y$ and projective inversion symmetry $P_I$ guarantees vanishing 2D Zak phase as well as quantized quadrupole moment $q_{xy}$ \cite{PhysRevB.96.245115}. In Fig. \ref{fig:7}(a), we calculate $q_{xy}$ versus $t_4$ with $t_1=t_2=2$, $t_3=1$ at $1/3$ filling, where an abrupt change from $0$ to $0.5$ can be observed at $t_4=1.86$. The change of $q_{xy}$ is associated with the vanishing of the Wannier band gap shown in Fig. \ref{fig:7}(b) as well as the change of nested Wilson loop $p^v$. The relationship between $q_{xy}$ and $p^v$ satisfies the typical characteristics of Type-I quadrupole topological insulator \cite{PhysRevResearch.2.033029}:
\begin{eqnarray} \label{eq33}
	q_{xy} = 2p_x^{v_y}p_y^{v_x}.
\end{eqnarray}
However, as shown in Fig. \ref{fig:6}(c), the critical point of the degenerate corner states is at $t_4=2$, which is inconsistent with $q_{xy}$. \emph{That is to say, although a nontrivial bulk topological phase can be defined in mirror-symmetric BBH$3$ model, the bulk-corner correspondence can not be established.} This mismatch between corner states and $q_{xy}$ indicates a fundamental difference between our new chiral corner states and the conventional symmetric-protected quadrupole corner states.

With the failure of bulk-corner correspondence, next we would like to establish the edge-corner correspondence. As shown in Fig. \ref{fig:7}(c), here we consider the $y$-open ribbon structure of the BBH$3$ model. From the band structure shown in Fig. \ref{fig:7}(d), three pairs of edge bands marked by different colors can be observed. We note that all the edge bands are two-fold degenerate due to the projective inversion symmetry. When $t_4$ passes the critical value, the edge gaps close and reopen while the parity of the edge states at the high-symmetry point flips, indicating the topological phase transition of the edge bands.

To exhibit the topological property of the edge band, we can calculate its polarization: 
\begin{eqnarray} \label{eq48}
	p_x^{edge}=\frac{i}{2\pi}\int dk_x \langle \psi_e^j(k_x) | \partial_{k_x} \psi_e^j(k_x) \rangle,
\end{eqnarray}
where $\psi_e^j(k_x)$ stands for the wavefunction of the $j$th edge band. Fig. \ref{fig:7}(e) exhibits $p_x^{edge}$ of the lowest two edge bands (marked by magenta lines in Fig. \ref{fig:7}(d)) versus $t_4$. When $t_4=2$, one can see an abrupt change from $p_x^{edge}=0$ to $p_x^{edge}=\pm0.5$, which is consistent with the appearance of the chiral corner states. In this way, the edge-corner correspondence can be established through the edge polarization $p^{edge}$.

We note that $p^{edge}$ is quantized only in mirror-symmetric systems. Therefore, for the general BBH$3$ model discussed in the next section, new method is needed to illustrate its higher-order topology.

\subsection{\label{sec:level1}Chiral Corner States in BBH3 Model without Mirror Symmetry}

\begin{figure}[t]
	\centering
	\includegraphics[width=1\linewidth]{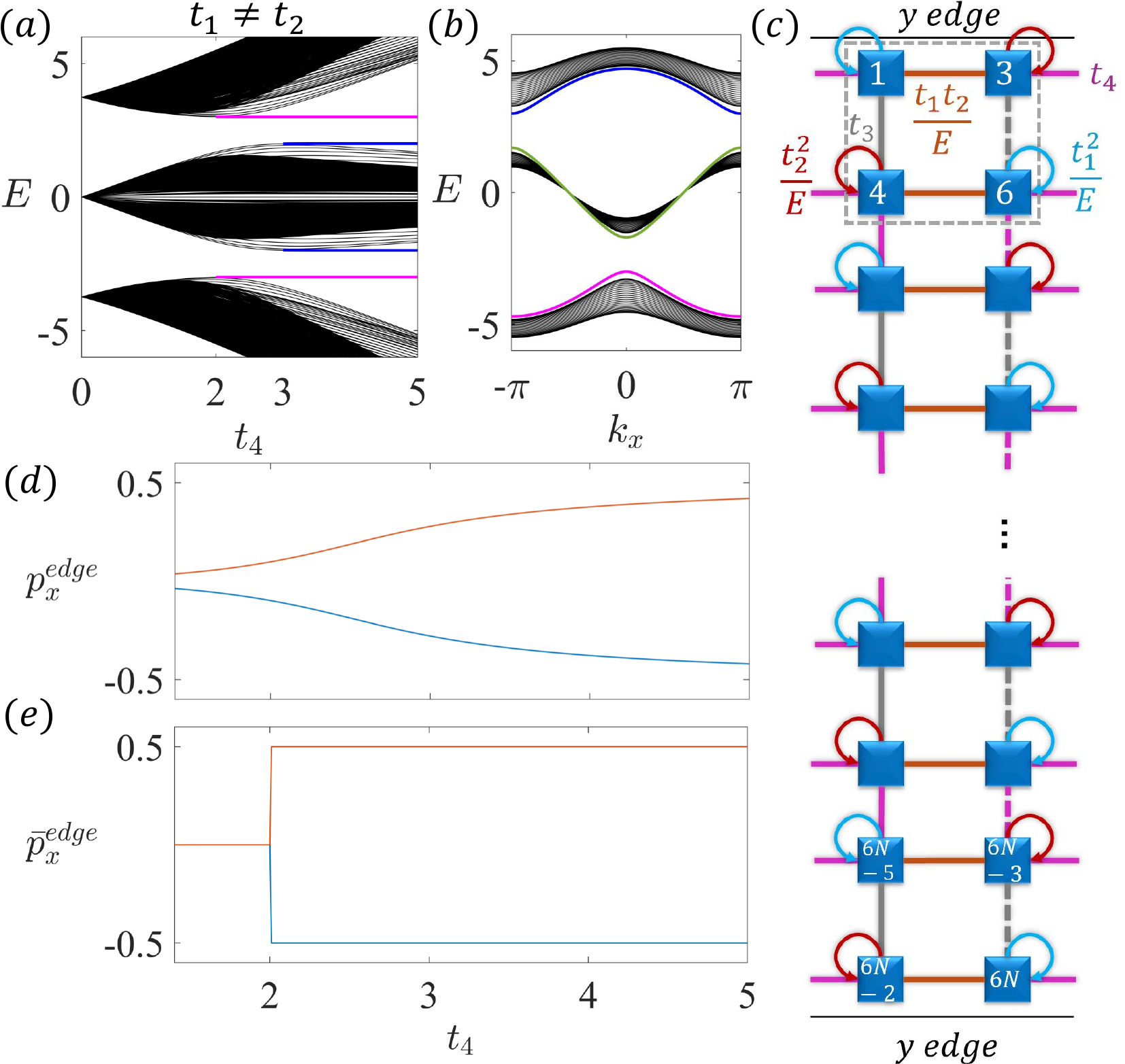}
	\caption{ (a) The band structure of the BBH$3$ model versus $t_4$ under OBC with $t_1=2$, $t_2=3$, $t_3=1$. (b) The band structure of the $y$-open system. (c) Schematic of the effective Hamiltonian of the $y$-open ribbon structure of the BBH$3$ model. (d) The edge polarization of the lowest two edge bands versus $t_4$. The edge polarization is no longer quantized due to the breaking of mirror symmetry. (e) The normalized edge polarization versus $t_4$.} \label{fig:8}
\end{figure}

By setting $t_1 \neq t_2$ in the BBH$3$ model, we can break the mirror symmetry in both $x$ and $y$ directions while preserving the spatial inversion symmetry. In this case, the band structure versus $t_4$ under OBC with $t_1=2$, $t_2=3$, $t_3=1$ is plotted in Fig. \ref{fig:8}(a). By breaking the mirror symmetry, the four-fold degenerate corner states in each gap split into two sets of two-fold degenerate corner states marked by different color with different energy and different critical point. Meanwhile, the emergence of the corner states is not accompanied by the closure of the band gaps, exhibiting similar characteristics of the asymmetric SSH$3$ model.

The split of the corner states can be understood through the redefined sublattices. Based on the Hamiltonian defined in Eq. \eqref{eq54}, Eq. \eqref{eq55} and the topological properties of 2D Rice-Mele model, we can deduce the energies and the critical points of chiral corner states as follows [Appendix \ref{D}]:
\begin{eqnarray} \label{eq56}
	\left\{\begin{matrix}
		E= \pm t_1, \quad t_4>\rm{max}\{ t_2,t_3 \}
		\\\\
		E= \pm t_2, \quad t_4>\rm{max}\{ t_1,t_3 \}
	\end{matrix}\right.,
\end{eqnarray}
which matches the band structure in Fig. \ref{fig:8}(a).

In terms of the edge-corner correspondence, as shown in Fig. \ref{fig:8}(b), three pairs of two-fold degenerate edge bands can still be observed. However, due to the breaking of mirror symmetry, the edge polarization in Fig. \ref{fig:8}(d) is no longer quantized. Meanwhile, we notice that the $y$-open ribbon structure of the BBH$3$ model illustrated in Fig. \ref{fig:7}(c) is effectively a 1D system, indicating that the methods introduced in Sec. \ref{III} might be utilized. Therefore, in Fig. \ref{fig:8}(c) we illustrate the effective Hamiltonian $\tilde{\mathcal{H}}_S(k_x,E)$ of the ribbon structure with $S=\{ | 6n-5 \rangle, | 6n-3 \rangle, | 6n-2 \rangle, | 6n \rangle \}$, where $n=1,...,N$ is the index of the unit cell in $y$ direction. $\tilde{\mathcal{H}}_S(k_x,E)$ show typical properties of Rice-Mele-like structure with $A=\{ | 6n-5 \rangle, | 6n \rangle \}$ and $B=\{ | 6n-3 \rangle, | 6n-2 \rangle \}$:
\begin{eqnarray} \label{eq82}
	\tilde{\mathcal{H}}_S(k_x,E) = \begin{bmatrix}
		\frac{t_1^2}{E}I_N & h(k_x,E)\\
		h^{\dagger}(k_x,E) & \frac{t_2^2}{E}I_N
	\end{bmatrix}.
\end{eqnarray}
Meanwhile, the effective Hamiltonian without onsite terms $\tilde{\mathcal{H}}^0_S(k_x)$ is mirror symmetric. Therefore, through the normalization of the reduced wavefunction in Eq. \eqref{eq29}, similar to the normalized $\bar{Z}$ defined in Sec. \ref{III}(B), we can calculate the normalized edge polarization:
\begin{eqnarray} \label{eq49}
	\bar{p}_x^{edge}=\frac{i}{2\pi}\int dk_x \langle \tilde{\psi}_e^j(k_x) | \partial_{k_x} \tilde{\psi}_e^j(k_x) \rangle.
\end{eqnarray}
The $\bar{p}_x^{edge}$ of the lowest two edge bands are shown in Fig. \ref{fig:8}(e). $\bar{p}_x^{edge}$ is quantized and an abrupt change from $\bar{p}^{edge}_x=0$ to $\bar{p}^{edge}_x=\pm 0.5$ can be observed at $t_4=2$, which coincides with the band structure in Fig. \ref{fig:8}(a). 

It is worth noting that the normalized $\bar{p}^{edge}_x$ also has the $\mathbb{Z}_2$ feature similar to the normalized $\bar{Z}$ defined in Sec. \ref{III}(B). That is to say, it cannot describe the case where even sets of corner states emerge simultaneously from the same edge band, such as the edge bands marked in green in Fig. \ref{fig:8}(b).

\begin{figure}[t]
	\centering
	\includegraphics[width=1\linewidth]{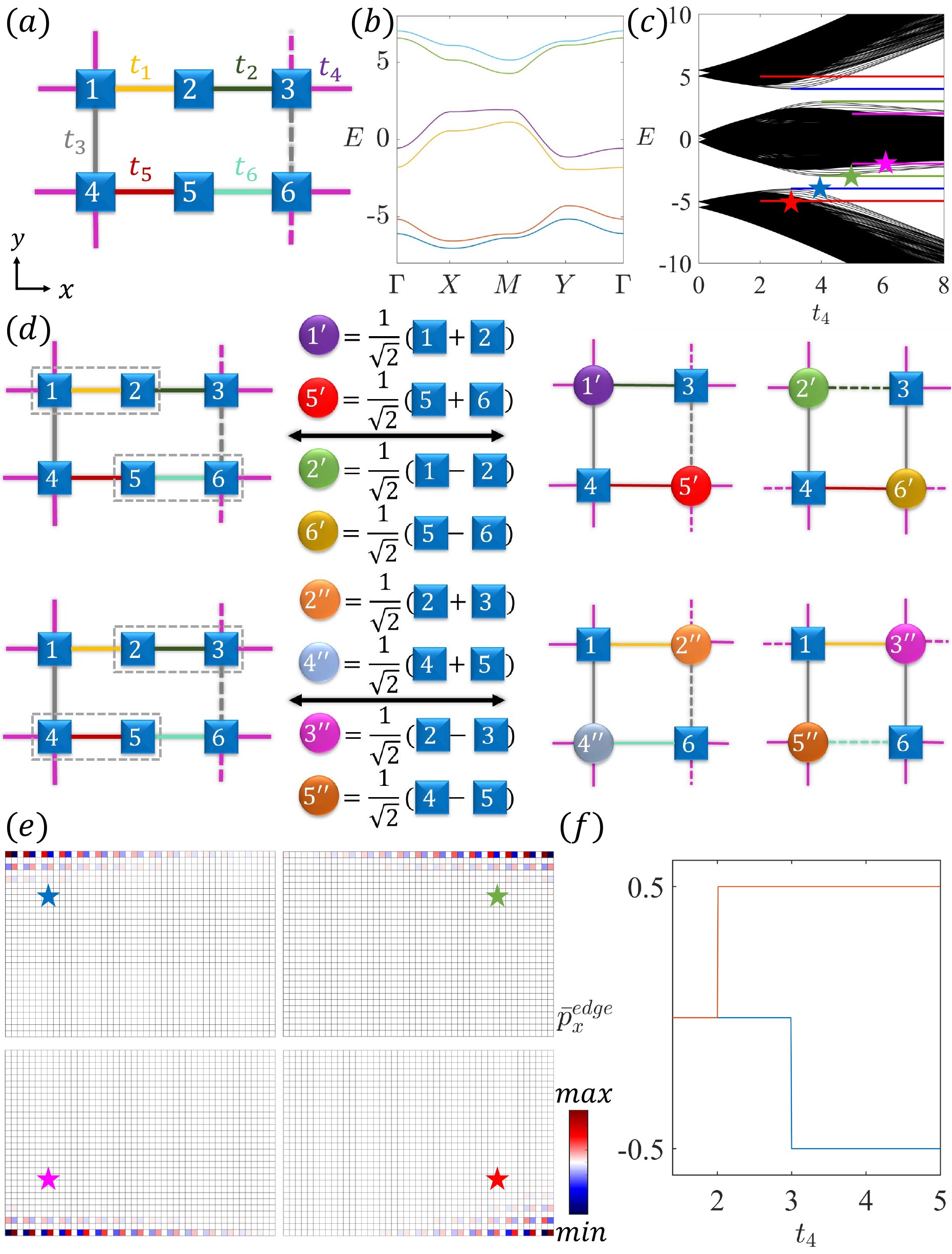}
	\caption{ (a) Schematic of the asymmetric BBH$3$ model without spatial inversion symmetry. (b) The bulk bands of the asymmetric BBH$3$ model. (c) The band structure versus $t_4$ under OBC with $t_1=2$, $t_2=5$, $t_3=1$, $t_5=4$, $t_6=3$. (d) Schematics of the asymmetric BBH$3$ model with two types of redefined sublattices. The system can no longer be decoupled into two 2D Rice-Mele models since the redefined sublattices on the opposite positions share different onsite energies. (e) Spatial distributions of the corner states marked by stars in (c). (f) The normalized edge polarization of the two lowest edge bands versus $t_4$. } \label{fig:9}
\end{figure}

\subsection{\label{sec:level1}Chiral Corner States in Asymmetric BBH3 Model without any Spatial Symmetry}

\begin{figure*}[t]
	\centering
	\includegraphics[width=1\linewidth]{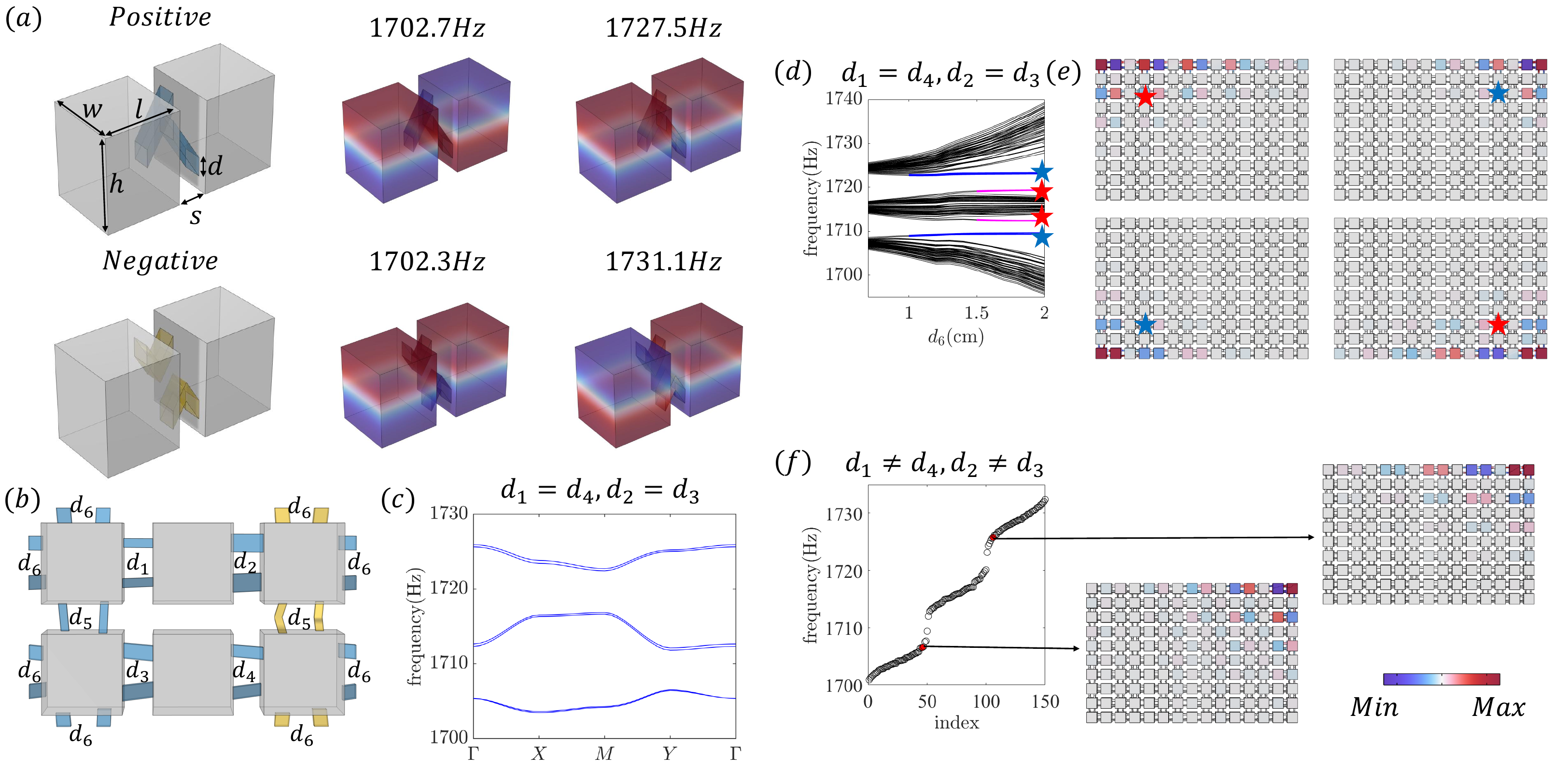}
	\caption{ (a) Acoustic design for the positive and negative hoppings. The size of the air cavity is set as $w=l=8$cm, $h=10$cm, $s=3$cm, $d=2$cm. (b) Schematic of the acoustic BBH$3$ model. (c) The bulk bands of the acoustic BBH$3$ model with $d_1=d_4=1$cm, $d_2=d_3=1.5$cm, $d_5=0.5$cm, $d_6=1$cm. (d) The band structure versus $d_6$ for a finite-sized ($5\times5$ unit cells) sample under OBC with $d_1=d_4=1$cm, $d_2=d_3=1.5$cm, $d_5=0.9$cm. The corner states are marked by red and blue lines. (e) The spatial distribution of the corner states in acoustics BBH$3$ model. The other four corner states can be obtained by inversion symmetry. (f) The band structure of the asymmetric BBH$3$ model under OBC with $d_1=0.9$cm, $d_2=1.8$cm, $d_3=1.5$cm, $d_4=1.2$cm, $d_5=0.8$cm, $d_6=1.4$cm. The spatial distribution of the chiral corner states marked by red points are plotted on the right side. } \label{fig:11}
\end{figure*}

To take a step further, in this section we would like to break the inversion symmetry in BBH$3$ model. As illustrated in Fig. \ref{fig:9}(a), we introduce the extra parameter $t_5$ and $t_6$ so that all the spatial symmetry is broken. The Hamiltonian of the asymmetric BBH$3$ model can be written as:
\begin{eqnarray} \label{eq83}
	\mathcal{H}_{ASBBH3}(k_x,k_y) = \begin{bmatrix}
		0 & t_1 & h_1' & h_2' & 0 & 0 \\
		t_1 & 0 & t_2 & 0 & 0 & 0 \\
		h_1'^* & t_2 & 0 & 0 & 0 & -h_2' \\
		h_2'^* & 0 & 0 & 0 & t_5 & h_1' \\
		0 & 0 & 0 & t_5 & 0 & t_6 \\
		0 & 0 & -h_2'^* & h_1'^* & t_6 & 0
	\end{bmatrix},
\end{eqnarray}
where $h_1' = t_4e^{ik_x}$, $h_2' = t_3+t_4e^{-ik_y}$. As a result, one can see in Fig. \ref{fig:9}(b) that the two-fold degeneracy of the bulk band is lifted. Surprisingly, as to the band structure under OBC shown in Fig. \ref{fig:9}(c), the eight in-gap corner states remain intact. Moreover, by turning the inter-cell hopping term $t_6$, the corner states marked by red and magenta lines move into the bulk continuum while their wavefunctions in Fig. \ref{fig:9}(e) are still localized at the corners, exhibiting the typical characteristics of the TBICs. 

In Fig. \ref{fig:9}(e), we also find that these corner states exhibit chiral characteristics. However, they can no longer be understood as the sub-symmetry protected chiral corner states with the redefined sublattices. When $t_1 \neq t_6$ and $t_2 \neq t_5$, as illustrated in Fig. \ref{fig:9}(d), the system after representation transformation can not be decomposed into two 2D Rice-Mele models since the redefined sublattices on the opposite positions, for example $| 1' \rangle$ and $| 5' \rangle$, $| 2' \rangle$ and $| 6' \rangle$, etc., share different effective onsite energies. Instead, these corner states can be explained as the sub-symmetry protected generalized chiral states \cite{wang2023sub,ni2019observation} as they only occupy one of the redefined sublattices. Based on the analysis in Appendix \ref{B}, we find that the energies and the critical points of these corner states can be written as:
\begin{eqnarray} \label{eq34}
	\left\{\begin{matrix}
		E= \pm t_1, \quad t_4>\rm{max}\{ t_2,t_3 \}
		\\\\
		E= \pm t_2, \quad t_4>\rm{max}\{ t_1,t_3 \}
		\\\\
		E= \pm t_6, \quad t_4>\rm{max}\{ t_5,t_3 \}
		\\\\
		E= \pm t_5, \quad t_4>\rm{max}\{ t_6,t_3 \}
	\end{matrix}\right.,
\end{eqnarray}
which is consistent with the band structure in Fig. \ref{fig:9}(c). As to the edge-corner correspondence, in Fig. \ref{fig:9}(f) we plot the normalized edge polarization $\bar{p}_x^{edge}$ of the lowest two edge bands. Two abrupt changes can be observed at $t_4 = t_1$ and $t_4 = t_6$ respectively, which exactly correspond to the critical points of the two lowest corner states.

\section{\label{sec:level1}TOPOLOGICAL CHIRAL CORNER STATES IN 2D ASYMMETRIC ACOUSTIC CRYSTALS} \label{V}

Mimic the experimental realization of the acoustic BBH model \cite{PhysRevLett.124.206601,xue2020observation}, in this section we propose an experimental design to realize the asymmetric BBH$3$ model and the corresponding chiral corner states. Here we utilize the air cavity working at the $P_z$-dipole modes to emulate the sublattices of the tight-binding model. As shown in Fig. \ref{fig:11}(a), the size of the air cavity is set to $w=l=8$cm, $h=10$cm, which presents a $P_z$-dipole mode with eigenfrequency $\omega_0=1716$Hz. By connecting two identical air cavities with tubes, the effective Hamiltonian of the composed system can be written as \cite{PhysRevLett.124.206601}:
\begin{eqnarray}
	H = \begin{bmatrix} \label{eq84}
		\omega_0 & \kappa \\
		\kappa & \omega_0
	\end{bmatrix},
\end{eqnarray}
where $\kappa$ stands for the coupling strength between the $P_z$ modes. For positive (negative) coupling, the $[1,1]^T$ ($[1,-1]^T$) eigenmode has a higher frequency. As shown in Fig. \ref{fig:11}(a), the positive and negative hoppings can be obtained through different types of connecting tubes between the air cavities, i.e. the straight tubes marked in blue and the twisted tubes marked in yellow. Meanwhile, the coupling strength can be tuned by the width of the connecting tube $d$. Based on these results, we design the acoustic BBH$3$ model, whose unit cell structure containing six air cavities is shown in Fig. \ref{fig:11}(b). 

We first consider the special case with $d_1=d_4$ and $d_2=d_3$, which corresponds to the BBH$3$ model discussed in Sec. \ref{IV}(D). The six dipole bulk bands are plotted in Fig. \ref{fig:11}(c). Similar to the tight-binding model, the bands are almost two-fold degenerate. In order to demonstrate the corner states, in Fig. \ref{fig:11}(d) we plot the band structure versus $d_4$ for a finite-sized ($5\times5$ unit cells) sample under OBC. As expected, two pairs of two-fold degenerate in-gap corner states marked by different colors can be observed with the critical points $d_6=d_1$ and $d_6=d_2$ respectively, which shares similar property with the band structure of the BBH$3$ model in Fig. \ref{fig:8}(a). Fig. \ref{fig:11}(e) shows the spatial distributions of the acoustic corner states. They also exhibit typical chiral characteristics in a way that only some of the air cavities in each unit cell are occupied. At last, we consider the general case with $d_1 \neq d_4$ and $d_2 \neq d_3$, which corresponds to the asymmetric BBH$3$ model discussed in Sec. \ref{IV}(E). In Fig. \ref{fig:11}(f), we find that two eigenstates marked by red circles are still chiral corner states even though they move into the bulk continuum. Therefore, all the topological chiral corner states obtained in Sec. \ref{IV} can be realized in the acoustic crystals proposed in this section.
	
\section{\label{sec:level1}CONCLUSION}

In summary, we systematically study the topological chiral boundary states in three different asymmetric models, i.e. 1D chain model, 1D ladder model and 2D model. Starting from the simplest SSH$m$ model, we find that all its possible edge states can be understood as the sub-symmetry protected chiral edge states with redefined sublattices. In terms of topological invariant, we define the quantized normalized Zak phase $\bar{Z}$ through the Rice-Mele-like effective Hamiltonian. The advantage of $\bar{Z}$ is that it can be easily extended to more complicated systems. As an example, we establish the bulk-edge correspondence in the asymmetric ladder SSH model. Furthermore, we introduce our methods into the 2D systems. We construct the 2D BBH$3$ model as a prototype to show our findings where new chiral corner states are found with redefined sublattices. These edge states are robust against any spatial symmetry, showing a fundamental difference with the spatial-symmetry protected quadrupole higher-order topology. The corner states can even move into the bulk continuous spectrum, exhibiting the characteristics of TBICs. Meanwhile, we also introduce the normalized $\bar{Z}$ into 2D and establish the edge-corner correspondence through the normalized edge polarization $\bar{p}^{edge}$. Finally, we propose an experimental realization of the acoustic BBH$3$ model and the chiral corner states are numerically observed. Our work establish the topological theory and the corresponding topological invariant for various asymmetric 1D and 2D asymmetric systems with chiral boundary states. By redefining sublattices, we find that models with entirely different structures might share the same topological origins. Utilizing our method, the discovery of the topological properties in many asymmetric systems can be expected.
	
\begin{acknowledgments}
		
This work is supported by National Natural Science Foundation of China (12174073).
		
\end{acknowledgments}
	
\appendix
	
\section{The Symmetry of the Normalized Reduced Wavefunction} \label{A}

In this appendix, we would like to show that the reduced wavefunction of the effective Hamiltonian satisfying Eq. \eqref{eq26} and Eq. \eqref{eq28} can reconstruct its symmetry through normalization. The generalized eigenequation in Eq. \eqref{eq27} can be decoupled into two parts:
\begin{eqnarray} \label{eq35}
	\left\{\begin{matrix}
		h(k,E)h^{\dagger}(k,E)\phi_A(k) = M(E)\phi_A(k)
		\\\\
		h^{\dagger}(k,E)h(k,E)\phi_B(k) = M(E)\phi_B(k).
	\end{matrix}\right.
\end{eqnarray}
with $M(E)=[E-v_1(E)][E-v_2(E)]$. Meanwhile, from Eq. \eqref{eq28} one can get:
\begin{eqnarray} \label{eq36}
	h(k,E) = h^{\dagger}(-k,E).
\end{eqnarray}
Since all the systems under consideration here have time-reversal symmetry, we have $\phi_A(k) = \phi_{A}^*(-k)$, $\phi_B(k) = \phi_{B}^*(-k)$. Therefore, the second equation in Eqs. \eqref{eq35} can be rewritten as:
\begin{eqnarray} \label{eq37}
	h(-k,E)h^{\dagger}(-k,E)\phi_B(k)=M(E)\phi_B(k),
\end{eqnarray}
which indicates that the difference between $\phi_B(k)$ and $\phi_A(-k)$ is up to a constant:
\begin{eqnarray} \label{eq38}
	\phi_B(k)=|C_k|e^{i\theta(k)}\phi_A(-k).
\end{eqnarray}
Following the normalization performed in Eq. \eqref{eq29}, we have $\tilde{\phi}_B(k)=e^{i\theta(k)}\tilde{\phi}_A(-k)$, $\tilde{\phi}_A(-k) = \tilde{\phi}^*_A(k)$, $\tilde{\phi}_B(-k) = \tilde{\phi}^*_B(k)$. It is easy to see that the normalized reduced wavefunction $\tilde{\psi}_S(k)$ is mirror symmetric:
\begin{eqnarray} \label{eq39}
	\begin{aligned}
		M_x\tilde{\psi}_S(k) = \begin{bmatrix}
			& I\\
			I & 
		\end{bmatrix}\begin{bmatrix}
			\tilde{\phi}_A(k)\\
			e^{i\theta(k)}\tilde{\phi}_A(-k)
		\end{bmatrix} = \\ e^{i\theta(k)}\begin{bmatrix}
			\tilde{\phi}_A(-k)\\
			e^{-i\theta(k)}\tilde{\phi}_A(k)
		\end{bmatrix} = e^{i\theta(k)}\tilde{\psi}_S(-k).
	\end{aligned}
\end{eqnarray}
Furthermore, as long as $\tilde{\psi}_S(k)$ is well-defined, it has a definite parity at the high-symmetry point $k_0=0,\pi$:
\begin{eqnarray} \label{eq40}
	M_x\tilde{\psi}_S(k_0) = \pm \tilde{\psi}_S(k_0).
\end{eqnarray}

\section{Detailed Derivation of the Sub-Symmetry Protected Chiral Edge States in Asymmetric Ladder SSH Model} \label{E}

In this appendix, we would like to give a detailed form of Eq. \eqref{eq53} and show that the edge states are sub-symmetry protected. $H'_{ladder}$ can be written as:
\begin{eqnarray} \label{eq77}
	H'_{ladder} = H_{ladder}^{(1)} + H_{ladder}^{(2)} + t_6P_{1',4'} - t_6P_{3',6'}
\end{eqnarray}
where 
\begin{eqnarray} \label{eq78}
	\begin{aligned}
		H_{ladder}^{(1)} = \sum_{n}\frac{t_1}{\sqrt{2}}c_{n,2}c^{\dagger}_{n,1'} + \frac{t_2}{\sqrt{2}}c_{n,2}c^{\dagger}_{n+1,1'} + \frac{t_5}{\sqrt{2}}c_{n,5}c^{\dagger}_{n,1'}
		\\
		 + \frac{t_5}{\sqrt{2}}c_{n,2}c^{\dagger}_{n,4'} + \frac{t_3}{\sqrt{2}}c_{n,5}c^{\dagger}_{n,4'} + \frac{t_4}{\sqrt{2}}c_{n+1,5}c^{\dagger}_{n,4'} + h.c.
	\end{aligned} \notag
	\\
	\begin{aligned}
		H_{ladder}^{(2)} = \sum_{n}\frac{t_1}{\sqrt{2}}c_{n,2}c^{\dagger}_{n,3'} + \frac{t_2}{\sqrt{2}}c_{n,2}c^{\dagger}_{n+1,3'} - \frac{t_5}{\sqrt{2}}c_{n,5}c^{\dagger}_{n,3'}
		\\
		+ \frac{t_5}{\sqrt{2}}c_{n,2}c^{\dagger}_{n,6'} - \frac{t_3}{\sqrt{2}}c_{n,5}c^{\dagger}_{n,6'} - \frac{t_4}{\sqrt{2}}c_{n+1,5}c^{\dagger}_{n,6'} + h.c.
	\end{aligned}
	\notag \\
\end{eqnarray}
and $P_{\alpha,\beta}$ stands for the projection operator over sublattice $| \alpha \rangle$,  $| \beta \rangle$. We note that the relationship between Eq. \eqref{eq53} and Eq. \eqref{eq77} reads as:
\begin{eqnarray} \label{eq85}
	\left\{\begin{matrix}
		H^{RM}_{\{1',2,5,4'\}} = H_{ladder}^{(1)} + t_6P_{1',4'}
		\\\\
		H^{RM}_{\{3',2,5,6'\}} = H_{ladder}^{(2)} - t_6P_{3',6'}
	\end{matrix}\right..
\end{eqnarray}
For $H_{ladder}^{(1)} + H_{ladder}^{(2)}$, the sub-symmetry over $| 1' \rangle$, $| 4' \rangle$ and $| 3' \rangle$, $| 6' \rangle$ are satisfied:
\begin{eqnarray} \label{eq79}
	\Gamma^{\dagger}_{(1)}[H_{ladder}^{(1)} + H_{ladder}^{(2)}]\Gamma_{(1)}P_{1',4'} = -[H_{ladder}^{(1)} + H_{ladder}^{(2)}]P_{1',4'} \notag \\
	\Gamma^{\dagger}_{(2)}[H_{ladder}^{(1)} + H_{ladder}^{(2)}]\Gamma_{(2)}P_{3',6'} = -[H_{ladder}^{(1)} + H_{ladder}^{(2)}]P_{3',6'} \notag \\
\end{eqnarray}
where $\Gamma_{(1)} = P_{1',4'} - P_{2,5}$, $\Gamma_{(2)} = P_{3',6'} - P_{2,5}$. Therefore, the zero-energy chiral edge states occupying sublattices $| 1' \rangle$, $| 4' \rangle$ and $| 3' \rangle$, $| 6' \rangle$ are sub-symmetry protected:
\begin{eqnarray} \label{eq80}
	\left\{\begin{matrix}
		\sum_{i=1}^2H_{ladder}^{(i)}| R_{1',4'} \rangle = \sum_{i=1}^2H_{ladder}^{(i)}| R_{3',6'} \rangle = 0
		\\\\
		P_{1',4'}| R_{1',4'} \rangle = | R_{1',4'} \rangle, \quad P_{3',6'}| R_{3',6'} \rangle = | R_{3',6'} \rangle
	\end{matrix}\right.
\end{eqnarray}
Together with Eq. \eqref{eq77}, we have:
\begin{eqnarray} \label{eq81}
	H'_{ladder}| R_{1',4'} \rangle = t_6 | R_{1',4'} \rangle, \quad H'_{ladder}| R_{3',6'} \rangle = -t_6| R_{3',6'} \rangle, \notag \\
\end{eqnarray}
i.e. chiral edge states can be found in asymmetric ladder SSH model at $E = \pm t_6$.

\section{The Detailed Topological Properties of the Chiral Corner States in BBH3 Model} \label{D}

In this section, we would like to deduce the energies and the critical points of the chiral corner states in BBH$3$ model. The two 2D Rice-Mele models defined in Eq. \eqref{eq54} can be written as:
\begin{eqnarray} \label{eq57}
	\mathcal{H}^{2DRM}_{\{1',3,4,5'\}} = \begin{bmatrix}
		t_1I_2 & \mathcal{H}'_1 \\
		\mathcal{H}'^{\dagger}_1 & 0
	\end{bmatrix}, \quad \mathcal{H}^{2DRM}_{\{2',3,4,6'\}} = \begin{bmatrix}
	-t_1I_2 & \mathcal{H}'_2 \\
	\mathcal{H}'^{\dagger}_2 & 0
	\end{bmatrix} \notag \\
\end{eqnarray}
with
\begin{eqnarray} \label{eq58}
	\left\{\begin{matrix}
		\mathcal{H}'_1 = \frac{1}{\sqrt{2}}\begin{bmatrix}
			t_2+t_4e^{ik_x} & t_3+t_4e^{-ik_y} \\
			-t_3-t_4e^{ik_y} & t_2+t_4e^{-ik_x}
		\end{bmatrix}
		\\\\
		\mathcal{H}'_2 = \frac{1}{\sqrt{2}}\begin{bmatrix}
			-t_2+t_4e^{ik_x} & t_3+t_4e^{-ik_y} \\
			t_3+t_4e^{ik_y} & t_2-t_4e^{-ik_x}
		\end{bmatrix}
	\end{matrix}\right..
\end{eqnarray}
Based on the topological property of 2D Rice-Mele model \cite{PhysRevB.110.104106}, for $\mathcal{H}^{2DRM}_{\{1',3,4,5'\}}$, when the inter-cell hopping exceeds the maximum of intra-cell hoppings, i.e. $t_4>\rm{max}\{ t_2, t_3 \}$, there exist chiral corner states occupying sublattice $| 1'  \rangle$, $| 5'  \rangle$ with $E = t_1$. For $\mathcal{H}^{2DRM}_{\{2',3,4,6'\}}$, chiral corner states occupying sublattice $| 2'  \rangle$, $| 6'  \rangle$ with $E = -t_1$ can be observed also when $t_4>\rm{max}\{ t_2, t_3 \}$. 

The two 2D Rice-Mele models defined in Eq. \eqref{eq55} can be written as:
\begin{eqnarray} \label{eq59}
	\mathcal{H}^{2DRM}_{\{1,2'',4'',6\}} = \begin{bmatrix}
		0 & \mathcal{H}''_1 \\
		\mathcal{H}''^{\dagger}_1 & t_2I_2
	\end{bmatrix}, \quad \mathcal{H}^{2DRM}_{\{1,3'',5'',6\}} = \begin{bmatrix}
		0 & \mathcal{H}''_2 \\
		\mathcal{H}''^{\dagger}_2 & -t_2I_2
	\end{bmatrix} \notag \\
\end{eqnarray}
with
\begin{eqnarray} \label{eq60}
	\left\{\begin{matrix}
		\mathcal{H}''_1 = \frac{1}{\sqrt{2}}\begin{bmatrix}
			t_1+t_4e^{ik_x} & t_3+t_4e^{-ik_y} \\
			-t_3-t_4e^{ik_y} & t_1+t_4e^{-ik_x}
		\end{bmatrix}
		\\\\
		\mathcal{H}''_2 = \frac{1}{\sqrt{2}}\begin{bmatrix}
			t_1-t_4e^{ik_x} & t_3+t_4e^{-ik_y} \\
			t_3+t_4e^{ik_y} & -t_1+t_4e^{-ik_x}
		\end{bmatrix}
	\end{matrix}\right..
\end{eqnarray}
Therefore, when $t_4>\rm{max}\{ t_1, t_3 \}$, there exist chiral corner states occupying sublattice $| 2''  \rangle$, $| 4''  \rangle$ ($| 3''  \rangle$, $| 5''  \rangle$) with $E = t_2$ ($E = -t_2$).
               
\section{The Generalized Chiral Corner States in Asymmetric BBH Model} \label{B}

\begin{figure}[t]
	\centering
	\includegraphics[width=1\linewidth]{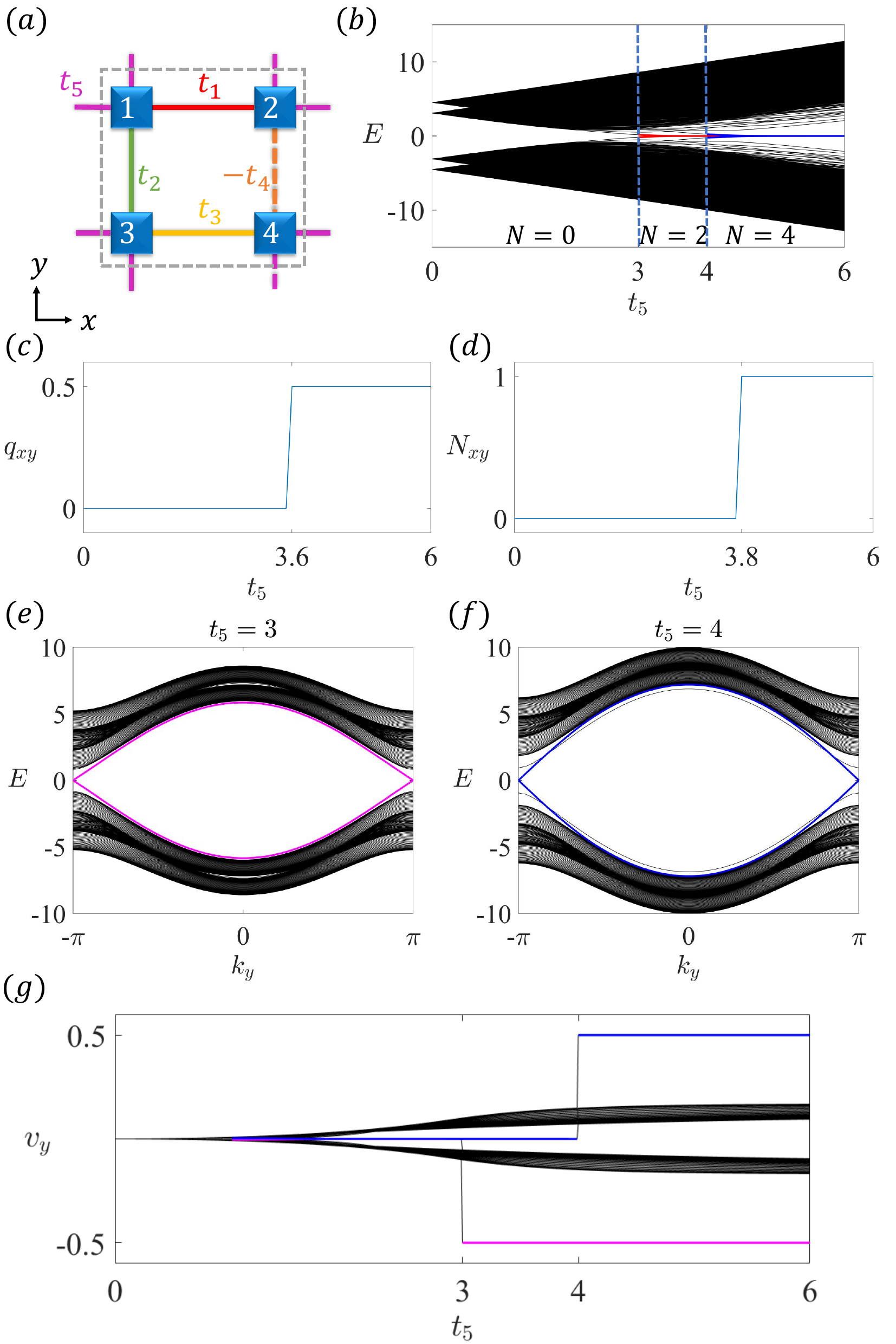}
	\caption{ (a) Schematic of the asymmetric BBH model. (b) The band structure versus $t_5$ under OBC with $t_1=1$, $t_2=3$, $t_3=2$, $t_4=4$. The spectrum is divided into three regions according to the number of zero-energy corner states. (c) The quadrupole moment $q_{xy}$ versus $t_5$. (d) The qdadrupole winding number $N_{xy}$ versus $t_5$. (e)-(f) The band structure of the $x$-open ribbon structure at (e) $t_5=3$ and $t_5=4$. (g) The Wannier band versus $t_5$ at half filling. The edge polarization $p_y^{edge}$ is marked by red and blue lines. } \label{fig:10}
\end{figure}

In this appendix, we would like to analyze the properties of the generalized chiral corner states in the asymmetric BBH model. As shown in Fig. \ref{fig:10}(a), the Hamiltonian of the asymmetric BBH model can be written as:
\begin{eqnarray} \label{eq41}
	\mathcal{H}_{ASBBH}(k_x,k_y) = \begin{bmatrix}
		0 & m_1 & m_2 & 0 \\
		m_1^* & 0 & 0 & m_3 \\
		m_2^* & 0 & 0 & m_4 \\
		0 & m_3^* & m_4^* & 0
	\end{bmatrix},
\end{eqnarray}
where $m_1=t_1+t_5e^{ik_x}$, $m_2=t_2+t_5e^{-ik_y}$, $m_3=-t_4-t_5e^{-ik_y}$, $m_4=t_3+t_5e^{ik_x}$. With the unequal intra-cell hopping terms, both the mirror symmetry and spatial inversion symmetry are broken. As to the non-spatial symmetry, the asymmetric BBH model has time-reversal symmetry and chiral symmetry. The band structure versus $t_5$ under OBC with $t_1=1$, $t_2=3$, $t_3=2$, $t_4=4$ is plotted in Fig. \ref{fig:10}(b), which can be divided into three regions according to the gap-closing points $t_5=3$ and $t_5=4$. In these three regions, the number of the zero-energy corner states is $N=0$, $N=2$ and $N=4$, respectively. This is quite different from the conventional BBH model \cite{doi:10.1126/science.aah6442} and the extended BBH model \cite{PhysRevLett.128.127601} where the number of the zero-energy corner states has to be a multiple of $4$. The difference is also verified through the bulk invariant. Although both the quadrupole moment $q_{xy}$ and the quadrupole winding number $N_{xy}$ are quantized due to chiral symmetry \cite{PhysRevLett.125.166801,PhysRevLett.128.127601}, neither of them is able to establish the bulk-corner correspondence.

The properties of the zero-energy corner states can be explained through the generalized chiral symmetry. For a system composed of $n$ sublattices in each unit cell, the generalized chiral symmetry is represented as \cite{ni2019observation,li2021zero}:
\begin{eqnarray} \label{eq42}
	\left\{\begin{matrix}
		\Gamma_n^{-1}\mathcal{H}_i\Gamma_n=\mathcal{H}_{i+1}, \quad i=0,...,n-2
		\\\\
		\sum_{j=0}^{n-1}\mathcal{H}_j=0
	\end{matrix}\right.,
\end{eqnarray}
where $\mathcal{H}_0$ is the system Hamiltonian and $\Gamma_n$ is the generalized chiral operator:
\begin{eqnarray} \label{eq43}
	\Gamma_n = \begin{bmatrix}
	1 & & & \\
	 & e^{i\frac{2\pi}{n}} & & \\
	 & & \ddots & \\
	 & & & e^{i\frac{2\pi(n-1)}{n}}
	\end{bmatrix}.
\end{eqnarray}
For the zero-energy corner state protected by the $\Gamma_n$, its wavefunction only occupies one of the sublattices \cite{ni2019observation}. Obviously, $\mathcal{H}_{ASBBH}(k_x,k_y)$ possesses the generalized chiral symmetry $\Gamma_4$. Therefore, the zero-energy corner states  $| C_\alpha \rangle$ occupying sublattice $\alpha=1,2,3,4$ should satisfy:
\begin{eqnarray} \label{eq44}
	\left\{\begin{matrix}
		\langle \beta | C_\alpha \rangle = \delta_{\alpha\beta}, \quad \beta=1,2,3,4
		\\\\
		\mathcal{H}_{ASBBH}| C_\alpha \rangle = 0
	\end{matrix}\right..
\end{eqnarray}
Therefore we have:
\begin{eqnarray} \label{eq45}
	\left\{\begin{matrix}
 		| C_1 \rangle : t_5>\rm{max}\{ t_1,t_2 \}
 		\\\\
 		| C_2 \rangle : t_5>\rm{max}\{ t_1,t_4 \}
 		\\\\
 		| C_3 \rangle : t_5>\rm{max}\{ t_2,t_3 \}
 		\\\\
 		| C_4 \rangle : t_5>\rm{max}\{ t_3,t_4 \}
 	\end{matrix}\right.
\end{eqnarray}
 which coincides with the band structure exhibited in Fig. \ref{fig:10}(b).
 
 The topological property of the asymmetric BBH model can be described through the edge polarization $p^{edge}$. In Fig. \ref{fig:10}(e)-(f), we plot the band structure of the $x$-open ribbon structure. Due to the breaking of spatial inversion symmetry, the degeneracy of the edge bands marked in blue and red is lifted. As a result, they share different topological transition points, which corresponds to the gap-closing points in Fig. \ref{fig:10}(b). It is worth noting that the edge bands are not isolated from the bulk bands, so the $p^{edge}$ can not be directly calculated by Eq. \eqref{eq48}. Instead, in Fig. \ref{fig:10}(g) we calculate the Wannier spectrum $v_y$ at half filling. The edge states in $v_y$ marked in color stands for $p^{edge}$, where the abrupt change takes place at $t_5=3$ and $t_5=4$ respectively, indicating the edge-corner correspondence is established. 
 
 As to the sub-symmetry property, the chiral corner states occupying sublattice $| \alpha \rangle$ is preserved as long as the following relationship is satisfied \cite{wang2023sub}:
 \begin{eqnarray} \label{eq46}
 	\sum_{j=0}^{n-1}\mathcal{H}_jP_\alpha=0,
 \end{eqnarray}
 where $\mathcal{H}_j$ is defined in Eqs. \eqref{eq42}, $P_\alpha$ is the projection operator $P_\alpha=| \alpha \rangle \langle \alpha |$. Meanwhile, the onsite term over sublattice $\alpha$ $V_\alpha$ only shifts the energy of the chiral corner states without affecting its spatial distribution. Therefore, for the asymmetric BBH$3$ model with redefined sublattices shown in Fig. \ref{fig:9}(d), the generalized chiral corner states occupying sublattices $| 1' \rangle$, $| 2' \rangle$, $| 5' \rangle$, $| 6' \rangle$, $| 2'' \rangle$, $| 3'' \rangle$, $| 4'' \rangle$, $| 5'' \rangle$ are preserved, which are exactly the eight corner states shown in Fig. \ref{fig:9}(c) and Fig. \ref{fig:9}(e).
	

\bibliography{reference}

\begin{thebibliography}{62}
\bibitem [{not()}]{note}%
  \BibitemOpen
  \href@noop {} {\bibinfo {title} {The analytic solution of the topological
  chiral edge states in the disordered {SSH} model utilizing the method in
  {Sec. II B} will be discussed in our other work, which will be published on
  other academic journal}}\BibitemShut {NoStop}%
\end{thebibliography}
	
\end{document}